%%%%%%%%%%%%%%%%%%%%%%%%%%%%%%%%%%%%%%%%%%%%%%%%%%%%%%%%%%%%%%%%%%%%%%%%%%%%%%
%%%%%%%%%%%%%%%%%%%%%%%%%%%bfklscale.tex%%%%%%%%%%%%%%%%%%%%%%%%%%%%%%%%%%%%%%
%%%%%%%%%%%%%%%%%%%%%%%%%%%%%%%%%%%%%%%%%%%%%%%%%%%%%%%%%%%%%%%%%%%%%%%%%%%%%%

%% site dependent options: 
%% \unredoffs and \redoffs define horizontal and vertical offsets 
%% respectively for unreduced and reduced modes. \speclscape defines
%% the \special{} call that sets printer to landscape (sideways) mode.
%% from standard set below, leave uncommented as appropriate or redefine
%
%%% next 400dpi
%\def\unredoffs{} \def\redoffs{\voffset=-.31truein\hoffset=-.48truein}
%\def\speclscape{\special{landscape}}
%
%%% apple lw
%\def\unredoffs{} \def\redoffs{\voffset=-.31truein\hoffset=-.59truein}
%\def\speclscape{\special{ps: landscape}}
%
%%% qms lasergrafix:
%\def\unredoffs{} \def\redoffs{\voffset=-.4truein\hoffset=.125truein}
%\def\speclscape{\special{qms: landscape}}
%
%%% saclay A4 paper:
\def\unredoffs{\hoffset-.14truein\voffset-.2truein} 
 
%\def\speclscape{\special{landscape}}
%
%---------------------------------------------------------------------%
%
\newbox\leftpage \newdimen\fullhsize \newdimen\hstitle \newdimen\hsbody
\tolerance=1000\hfuzz=2pt
\catcode`\@=11 % This allows us to modify PLAIN macros.
%\def\bigans{b }
%\message{ big or little (b/l)? }\read-1 to\answ
%
%\ifx\answ\bigans\message{(This will come out unreduced.}
\magnification=1095\unredoffs\baselineskip=16pt plus 2pt minus 1pt
\hsbody=\hsize \hstitle=\hsize %take default values for unreduced format
%
%\else\message{(This will be reduced.} \let\l@r=L
%\magnification=800\baselineskip=16pt plus 1pt minus 0.5pt \vsize=7truein
%\redoffs \hstitle=8truein\hsbody=4.75truein\fullhsize=10truein\hsize=\hsbody
%
%\output={\ifnum\pageno=0 %%% This is the HUTP version
%  \shipout\vbox{\speclscape{\hsize\fullhsize\makeheadline}
%    \hbox to \fullhsize{\hfill\pagebody\hfill}}\advancepageno}
%  \else
% \almostshipout{\leftline{\vbox{\pagebody\makefootline}}}\advancepageno 
%  \fi}
%\def\almostshipout#1{%\if L\l@r \count1=1 \message{[\the\count0.\the\count1]}
%      \global\setbox\leftpage=#1 \global\let\l@r=R
% \else 
%\count1=2
%  \shipout\vbox{\speclscape{\hsize\fullhsize\makeheadline}
%      \hbox to\fullhsize{\box\leftpage\hfil#1}}  \global\let\l@r=L\fi}
%\fi
%---------------------------------------------------------------------
%
\newcount\yearltd\yearltd=\year\advance\yearltd by -1900

%
% 	restores pagenumbers
%
%       use following instead of \Date on the preliminary draft, 
%       puts date/time on each page in big mode, writes labels in margins

\def\draftmode{\message{ DRAFTMODE }\def\draftdate{{\rm preliminary draft:
\number\month/\number\day/\number\yearltd\ \ \hourmin}}%
\headline={\hfil\draftdate}\writelabels\baselineskip=16pt plus 2pt minus 2pt
 {\count255=\time\divide\count255 by 60 \xdef\hourmin{\number\count255}
  \multiply\count255 by-60\advance\count255 by\time
  \xdef\hourmin{\hourmin:\ifnum\count255<10 0\fi\the\count255}}}
%       use \nolabels to get rid of eqn, ref, and fig labels in draft mode
\def\nolabels{\def\wrlabeL##1{}\def\eqlabeL##1{}\def\reflabeL##1{}}
\def\writelabels{\def\wrlabeL##1{\leavevmode\vadjust{\rlap{\smash%
{\line{{\escapechar=` \hfill\rlap{\sevenrm\hskip.03in\string##1}}}}}}}%
\def\eqlabeL##1{{\escapechar-1\rlap{\sevenrm\hskip.05in\string##1}}}%
\def\reflabeL##1{\noexpand\llap{\noexpand\sevenrm\string\string\string##1}}}
\nolabels
%
% tagged sec numbers
\global\newcount\secno \global\secno=0
\global\newcount\meqno \global\meqno=1
\def\newsec#1{\global\advance\secno by1\message{(\the\secno. #1)}
%\ifx\answ\bigans \vfill\eject \else \bigbreak\bigskip \fi  %if desired
\global\subsecno=0\eqnres@t\noindent{\bf\the\secno. #1}
\writetoca{{\secsym} {#1}}\par\nobreak\medskip\nobreak}
\def\eqnres@t{\xdef\secsym{\the\secno.}\global\meqno=1\bigbreak\bigskip}
\def\sequentialequations{\def\eqnres@t{\bigbreak}}\xdef\secsym{}
\global\newcount\subsecno \global\subsecno=0
\def\subsec#1{\global\advance\subsecno by1\message{(\secsym\the\subsecno. #1)}
\ifnum\lastpenalty>9000\else\bigbreak\fi
\noindent{\bf\secsym\the\subsecno. #1}\writetoca{\string\quad 
{\secsym\the\subsecno.} {#1}}\par\nobreak\medskip\nobreak}
\def\appendix#1#2{\global\meqno=1\global\subsecno=0\xdef\secsym{\hbox{#1.}}
\bigbreak\bigskip\noindent{\bf Appendix #1. #2}\message{(#1. #2)}
\writetoca{Appendix {#1.} {#2}}\par\nobreak\medskip\nobreak}
%
%       \eqn\label{a+b=c}	gives displayed equation, numbered
%				consecutively within sections.
%     \eqnn and \eqna define labels in advance (of eqalign?)
%
\def\eqnn#1{\xdef #1{(\secsym\the\meqno)}\writedef{#1\leftbracket#1}%
\global\advance\meqno by1\wrlabeL#1}
\def\eqna#1{\xdef #1##1{\hbox{$(\secsym\the\meqno##1)$}}
\writedef{#1\numbersign1\leftbracket#1{\numbersign1}}%
\global\advance\meqno by1\wrlabeL{#1$\{\}$}}
\def\eqn#1#2{\xdef #1{(\secsym\the\meqno)}\writedef{#1\leftbracket#1}%
\global\advance\meqno by1$$#2\eqno#1\eqlabeL#1$$}
%
%			 footnotes
\newskip\footskip\footskip14pt plus 1pt minus 1pt %sets footnote baselineskip
\def\footnotefont{\ninepoint}\def\f@t#1{\footnotefont #1\@foot}
\def\f@@t{\baselineskip\footskip\bgroup\footnotefont\aftergroup\@foot\let\next}
\setbox\strutbox=\hbox{\vrule height9.5pt depth4.5pt width0pt}
\global\newcount\ftno \global\ftno=0
\def\foot{\global\advance\ftno by1\footnote{$^{\the\ftno}$}}
%
%say \footend to put footnotes at end
%will cause problems if \ref used inside \foot, instead use \nref before
\newwrite\ftfile   
\def\footend{\def\foot{\global\advance\ftno by1\chardef\wfile=\ftfile
$^{\the\ftno}$\ifnum\ftno=1\immediate\openout\ftfile=foots.tmp\fi%
\immediate\write\ftfile{\noexpand\smallskip%
\noexpand\item{f\the\ftno:\ }\pctsign}\findarg}%
\def\footatend{\vfill\eject\immediate\closeout\ftfile{\parindent=20pt
\centerline{\bf Footnotes}\nobreak\bigskip\input foots.tmp }}}
\def\footatend{}
%
%     \ref\label{text}
% generates a number, assigns it to \label, generates an entry.
% To list the refs on a separate page,  \listrefs
%
\global\newcount\refno \global\refno=1
\newwrite\rfile
\def\ref{[\the\refno]\nref}
\def\nref#1{\xdef#1{[\the\refno]}\writedef{#1\leftbracket#1}%
\ifnum\refno=1\immediate\openout\rfile=refs.tmp\fi
\global\advance\refno by1\chardef\wfile=\rfile\immediate
\write\rfile{\noexpand\item{#1\ }\reflabeL{#1\hskip.31in}\pctsign}\findarg}
%	horrible hack to sidestep tex \write limitation
\def\findarg#1#{\begingroup\obeylines\newlinechar=`\^^M\pass@rg}
{\obeylines\gdef\pass@rg#1{\writ@line\relax #1^^M\hbox{}^^M}%
\gdef\writ@line#1^^M{\expandafter\toks0\expandafter{\striprel@x #1}%
\edef\next{\the\toks0}\ifx\next\em@rk\let\next=\endgroup\else\ifx\next\empty%
\else\immediate\write\wfile{\the\toks0}\fi\let\next=\writ@line\fi\next\relax}}
\def\striprel@x#1{} \def\em@rk{\hbox{}} 
\def\lref{\begingroup\obeylines\lr@f}
\def\lr@f#1#2{\gdef#1{\ref#1{#2}}\endgroup\unskip}
\def\semi{;\hfil\break}
\def\addref#1{\immediate\write\rfile{\noexpand\item{}#1}} %now unnecessary
\def\startrefs#1{\immediate\openout\rfile=refs.tmp\refno=#1}
\def\xref{\expandafter\xr@f}\def\xr@f[#1]{#1}
\def\refs#1{\count255=1[\r@fs #1{\hbox{}}]}
\def\r@fs#1{\ifx\und@fined#1\message{reflabel \string#1 is undefined.}%
\nref#1{need to supply reference \string#1.}\fi%
\vphantom{\hphantom{#1}}\edef\next{#1}\ifx\next\em@rk\def\next{}%
\else\ifx\next#1\ifodd\count255\relax\xref#1\count255=0\fi%
\else#1\count255=1\fi\let\next=\r@fs\fi\next}
%

%
% this is ugly, but moore insists
\newwrite\ffile\global\newcount\figno \global\figno=1
\def\fig{fig.~\the\figno\nfig}
\def\nfig#1{\xdef#1{fig.~\the\figno}%
\writedef{#1\leftbracket fig.\noexpand~\the\figno}%
\ifnum\figno=1\immediate\openout\ffile=figs.tmp\fi\chardef\wfile=\ffile%
\immediate\write\ffile{\noexpand\medskip\noexpand\item{Fig.\ \the\figno. }
\reflabeL{#1\hskip.55in}\pctsign}\global\advance\figno by1\findarg}
\def\xfig{\expandafter\xf@g}\def\xf@g fig.\penalty\@M\ {}
\def\figs#1{figs.~\f@gs #1{\hbox{}}}
\def\f@gs#1{\edef\next{#1}\ifx\next\em@rk\def\next{}\else
\ifx\next#1\xfig #1\else#1\fi\let\next=\f@gs\fi\next}
\newwrite\lfile
{\escapechar-1\xdef\pctsign{\string\%}\xdef\leftbracket{\string\{}
\xdef\rightbracket{\string\}}\xdef\numbersign{\string\#}}

\def\writestop{\def\writestoppt{\immediate\write\lfile{\string\pageno%
\the\pageno\string\startrefs\leftbracket\the\refno\rightbracket%
\string\def\string\secsym\leftbracket\secsym\rightbracket%
\string\secno\the\secno\string\meqno\the\meqno}\immediate\closeout\lfile}}
\def\writestoppt{}\def\writedef#1{}
\def\seclab#1{\xdef #1{\the\secno}\writedef{#1\leftbracket#1}\wrlabeL{#1=#1}}
\def\subseclab#1{\xdef #1{\secsym\the\subsecno}%
\writedef{#1\leftbracket#1}\wrlabeL{#1=#1}}
\newwrite\tfile \def\writetoca#1{}
\def\leaderfill{\leaders\hbox to 1em{\hss.\hss}\hfill}
%	use this to write file with table of contents
\def\writetoc{\immediate\openout\tfile=toc.tmp 
   \def\writetoca##1{{\edef\next{\write\tfile{\noindent ##1 
   \string\leaderfill {\noexpand\number\pageno} \par}}\next}}}
%       and this lists table of contents on second pass

%
\catcode`\@=12 % at signs are no longer letters
%
%	Unpleasantness in calling in abstract and title fonts
\edef\tfontsize{\ifx\answ\bigans scaled\magstep3\else scaled\magstep4\fi}
 \tfontsize  \tfontsize
 \tfontsize \font\titlei=cmmi10 \tfontsize
\font\titleis=cmmi7 \tfontsize \font\titleiss=cmmi5 \tfontsize
\font\titlesy=cmsy10 \tfontsize \font\titlesys=cmsy7 \tfontsize
\font\titlesyss=cmsy5 \tfontsize  \tfontsize
\skewchar\titlei='177 \skewchar\titleis='177 \skewchar\titleiss='177
\skewchar\titlesy='60 \skewchar\titlesys='60 \skewchar\titlesyss='60
 \ifx\answ\bigans\else scaled\magstep1\fi
\ifx\answ\bigans\else

 \font\absi=cmmi10 scaled\magstep1
\font\absis=cmmi7 scaled\magstep1 \font\absiss=cmmi5 scaled\magstep1
\font\abssy=cmsy10 scaled\magstep1 \font\abssys=cmsy7 scaled\magstep1
\font\abssyss=cmsy5 scaled\magstep1 
\skewchar\absi='177 \skewchar\absis='177 \skewchar\absiss='177
\skewchar\abssy='60 \skewchar\abssys='60 \skewchar\abssyss='60
\fi
\font\ninerm=cmr9 \font\sixrm=cmr6 \font\ninei=cmmi9 \font\sixi=cmmi6 
\font\ninesy=cmsy9 \font\sixsy=cmsy6 \font\ninebf=cmbx9 
\font\nineit=cmti9 \font\ninesl=cmsl9 \skewchar\ninei='177
\skewchar\sixi='177 \skewchar\ninesy='60 \skewchar\sixsy='60 
\def\ninepoint{\def\rm{\fam0\ninerm}% switch to footnote font
\textfont0=\ninerm \scriptfont0=\sixrm \scriptscriptfont0=\fiverm
\textfont1=\ninei \scriptfont1=\sixi \scriptscriptfont1=\fivei
\textfont2=\ninesy \scriptfont2=\sixsy \scriptscriptfont2=\fivesy
\textfont\itfam=\ninei \def\it{\fam\itfam\nineit}\def\sl{\fam\slfam\ninesl}%
\textfont\bffam=\ninebf \def\bf{\fam\bffam\ninebf}\rm} 
%
%---------------------------------------------------------------------
%

\hyphenation{anom-aly anom-alies coun-ter-term coun-ter-terms}
\def\inv{^{\raise.15ex\hbox{${\scriptscriptstyle -}$}\kern-.05em 1}}

\def\Dsl{\,\raise.15ex\hbox{/}\mkern-13.5mu D} %this one can be subscripted
\def\dsl{\raise.15ex\hbox{/}\kern-.57em\partial}

 %pound sterling
\def\lspace{\ifx\answ\bigans{}\else\qquad\fi}
\def\lbspace{\ifx\answ\bigans{}\else\hskip-.2in\fi} % $$\lbspace...$$
\def\boxeqn#1{\vcenter{\vbox{\hrule\hbox{\vrule\kern3pt\vbox{\kern3pt
	\hbox{${\displaystyle #1}$}\kern3pt}\kern3pt\vrule}\hrule}}}
\def\mbox#1#2{\vcenter{\hrule \hbox{\vrule height#2in
		\kern#1in \vrule} \hrule}}  %e.g. \mbox{.1}{.1}
%	matters of taste
%\def\tilde{\widetilde} \def\bar{\overline} \def\hat{\widehat}
%
% some sample definitions
  %     curly letters

\def\darr#1{\raise1.5ex\hbox{$\leftrightarrow$}\mkern-16.5mu #1}
 %pound sterling

\def\half{{\textstyle{1\over2}}} %puts a small half in a displayed eqn
\def\fourninths{{\textstyle{4\over9}}} %ditto 4/9
 %ditto 3/2
\def\roughly#1{\raise.3ex\hbox{$#1$\kern-.75em\lower1ex\hbox{$\sim$}}}

\def\p2inf{\mathrel{\mathop{\sim}\limits_{\scriptscriptstyle
{p^2 \rightarrow \infty }}}}
\def\kap2inf{\mathrel{\mathop{\sim}\limits_{\scriptscriptstyle
{\kappa \rightarrow \infty }}}}
\def\x2inf{\mathrel{\mathop{\sim}\limits_{\scriptscriptstyle
{x \rightarrow \infty }}}}
\def\Lam2inf{\mathrel{\mathop{\sim}\limits_{\scriptscriptstyle
{\Lambda \rightarrow \infty }}}}
\def\frac#1#2{{{#1}\over {#2}}}
\def\half{\hbox{${1\over 2}$}}

\def\Gev{{\rm GeV}}

\def\lsim{\mathrel{mathpalette\@v1000ersim<}}
\def\gsim{\mathrel{mathpalette\@versim>}}

\catcode`@=11 %This allows us to modify plain macros
\def\slash#1{\mathord{\mathpalette\c@ncel#1}}
 \def\c@ncel#1#2{\ooalign{$\hfil#1\mkern1mu/\hfil$\crcr$#1#2$}}
\def\lsim{\mathrel{\mathpalette\@versim<}}
\def\gsim{\mathrel{\mathpalette\@versim>}}
 \def\@versim#1#2{\lower0.2ex\vbox{\baselineskip\z@skip\lineskip\z@skip
       \lineskiplimit\z@\ialign{$\m@th#1\hfil##$\crcr#2\crcr\sim\crcr}}}
\catcode`@=12 %at signs are no longer letters

\def\PR{{\it Phys.~Rev.~}}
\def\PRL{{\it Phys.~Rev.~Lett.~}}
\def\NP{{\it Nucl.~Phys.~}}
\def\PL{{\it Phys.~Lett.~}}
\def\PRep{{\it Phys.~Rep.~}}

\def\SJNP{{\it Sov.~Jour.~Nucl.~Phys.~}}
\def\ZP{{\it Zeit.~Phys.~}}

\def\vol#1{{\bf #1}}
\def\vyp#1#2#3{\vol{#1} (#2) #3}

\def\Asl{\raise.15ex\hbox{/}\mkern-11.5mu A}
\def\psl{\lower.12ex\hbox{/}\mkern-9.5mu p}
\def\qsl{\lower.12ex\hbox{/}\mkern-9.5mu q}
\def\rsl{\lower.03ex\hbox{/}\mkern-9.5mu r}
\def\ksl{\raise.06ex\hbox{/}\mkern-9.5mu k}

%%%%%%%%%%%%%%%%%%%%%%%%%%%%%%%%%%%%%%%%%%%%%%%%%%%%%%%%%%%%%%%%%%%%%%%%%%%%%%%

\pageno=0\nopagenumbers\tolerance=10000\hfuzz=5pt
\line{\hfill OUTP-9902P}
\vskip 36pt
\centerline{\bf NLO BFKL Equation, Running Coupling}
\vskip 6pt
\centerline{\bf and Renormalization Scales.}
\vskip 36pt
\centerline{Robert~S.~Thorne}
\vskip 12pt
\centerline{\it Jesus College and Theoretical Physics,}
\centerline{\it University of Oxford, Oxford, Oxon., OX1 3DW, U.K.}
\vskip 0.9in
{\narrower\baselineskip 10pt
\centerline{\bf Abstract}
\medskip
I examine the solution of the BFKL equation with NLO corrections
relevant for deep inelastic scattering. Particular
emphasis is placed on the part played by the running of the coupling. It is
shown that the solution factorizes into a part describing the evolution in 
$Q^2$, and a constant part describing the input distribution. The latter is 
infrared dominated, being described by a coupling which grows as $x$ 
decreases, and thus being contaminated by infrared renormalons. 
Hence, for this part we agree with previous assertions that predictive
power breaks down for small enough $x$ at any $Q^2$. However, the former is
ultraviolet dominated, being described by a coupling which falls like 
$1/(\ln(Q^2/\Lambda^2) + A(\bar\alpha_s(Q^2)\ln(1/x))^{\half})$ with
decreasing $x$, and thus is perturbatively calculable at all $x$. Therefore,
although the BFKL equation is unable to predict the input for a structure
function for small $x$, it is able to predict its evolution in $Q^2$, 
as we would expect
from the factorization theory. The evolution at small $x$ has no true 
powerlike
behaviour due to the fall of the coupling, but does have significant
differences from that predicted from a standard NLO in $\alpha_s$
treatment. Application of the resummed splitting functions with the 
appropriate coupling constant to an analysis of data, i.e. a global
fit, is very successful.}
   
\vskip 0.7in
\line{\hfill}
\line{January 1999\hfill}
\vfill\eject
\footline={\hss\tenrm\folio\hss}

%%%%%%%%%%%%%%%%%%%%%%%%%%%%%%%%%%%%%%%%%%%%%%%%%%%%%%%%%%%%%%%%%%%%%%%%%%%%%%%

%%%%%%%%%%%%%%%%%%%%%%%%%%%%%%%%%%%%%%%%%%%%%%%%%%%%%%%%%%%%%%%%%%%%%%%%%%%%%%%

\newsec{Introduction.}

There has recently been a great deal of interest in the solution to the 
BFKL equation \ref\BFKL{L.N. Lipatov, \SJNP \vyp{23}{1976}{338}\semi
E.A. Kuraev, L.N. Lipatov and V.S. Fadin, {\it Sov.~Jour. JETP} 
\vyp{45}{1977}{199}\semi
Ya. Balitskii and L.N. Lipatov, \SJNP \vyp{28}{1978}{6}.}, 
triggered by the calculation of the NLO correction to the kernel 
\ref\NLOBFKLlf{V.S. Fadin and L.N. Lipatov, \PL \vyp{B429}{1998}{127}, and
references therein.}\ref\NLOBFKLcc{G. Camici and M. Ciafaloni, 
\PL \vyp{B430}{1998}{349}, and references therein.} 
and the apparent result that this leads to a huge correction to the LO result.
A number of subsequent papers have examined the solutions to this equation
and/or its consequences
\ref\bluvogt{J. Bl\"umlein, V. Ravindran, W.L.
van Neerven and  A. Vogt, Proceedings of DIS 98, 
Brussels, April 1998, p. 211, {\tt hep-ph/9806368}.}%
\nref\ballforte{R.D. Ball 
and S. Forte Proceedings of DIS 98, Brussels, April 1998, p. 770,
{\tt hep-ph/9805315}.}%
\nref\ross{D.A. Ross, \PL \vyp{B431}{1998}{161}.}%
\nref\mplusk{Yu. V. Kovchegov 
and A.H. Mueller, \PL \vyp{B439}{1998}{423}.}%
\nref\levin{E.M. Levin, 
Tel Aviv University, Report No. TAUP 2501-98, 
{\tt hep-ph/9806228}.}%
--\ref\armesto{N. Armesto, J. Bartels and M.A. Braun, \PL 
\vyp{B442}{1998}{459}.}
drawing a variety of conclusions (dominant negative NLO 
anomalous dimensions, oscillatory behaviour, non-Regge terms, 
instability, breakdown of perturbation theory), 
most being rather pessimistic. This has
prompted work on ways to at least estimate contributions to the kernel at 
even higher orders, and obtain perturbative stability via a summation 
\ref\gavin{G. Salam, JHEP
\vyp{9807}{1998}{19}.}\ref\ciafpcol{M. Ciafaloni 
and D. Colferai, {\tt hep-ph/9812366}.}. 

I will take the point of view that the most significant result
of the NLO kernel is that it indicates very strongly how the coupling 
constant should run in the BFKL equation, i.e. that the scale in the 
coupling should be chosen to be the transverse momentum at the top of the 
gluon ladder $k^2$. Making this choice \ref\glr{L.V. Gribov, E.M. Levin and 
M.G. Ryskin, \PRep \vyp{100}{1983}{1}.} I follow many previous authors in
examining how this choice affects the solution to the LO equation
\ref\liprun{L.N. Lipatov, {\it Sov. Phys. JETP} \vyp{63}{1986}{904}.}%
\nref\jan{J. Kwiecinski, \ZP \vyp{C29}{1985}{561}.}%
\nref\janpcoll{J.C. Collins
and J. Kwiecinski, \NP \vyp{B316}{1989}{307}.}%
\nref\hancock{R.E. Hancock and 
D.A. Ross, \NP \vyp{B383}{1992}{575}; \NP \vyp{B394}{1993}{200}.}%
\nref\nik{N.N. Nikolaev and B.G. 
Zakharov, \PL \vyp{B327}{1994}{157}.}%
\nref\levrun{E.M. Levin, \NP \vyp{B453}{1995}{303}.}%
\nref\Andersson{B. Andersson, G. Gustafson and 
H. Kharraziha, \PR \vyp{D57}{1998}{5543}.}%
\nref\haakman{L.P.A. Haakman, O.V. 
Kancheli, and J.H. Koch, \PL \vyp{B391}{1997}{157}; \NP 
\vyp{B518}{1998}{275}.}%
--\ref\ciafrun{G. Camici and M. Ciafaloni, \PL
\vyp{B395}{1997}{118}.}. 
I find that 
at leading twist the solution factorizes into a part dependent of the 
input to the equation, but independent of the scale $k^2$, and a part 
independent of the input which governs the evolution in $k^2$ 
\jan\janpcoll\ciafrun. The former is 
disastrously contaminated by the diffusion \ref\diff{J.Bartels and
H. Lotter, \PL \vyp{B309}{1993}{400}\semi
J. Bartels, H. Lotter and M. Vogt
\PL \vyp{B373}{1996}{215}.} 
into the infrared, and without a 
low $k^2$ regularization is indeterminate due to the presence of 
infrared renormalons giving behaviour 
$\sim \exp(-n\beta_0(\ln(Q_0^2/\Lambda^2))^{3}/(A^2\ln(1/x)))$, where $Q_0^2$
is the scale of the input to the equation, $n$ is an integer, and 
$A\sim 4$. This is entirely consistent 
with Mueller's result \ref\mueller{A.H. Mueller, \PL 
\vyp{B396}{1997}{251}.}\mplusk\ on the range of applicability of the 
BFKL equation.
It renders the NLO correction to the kernel which is not associated
with running
of the coupling rather unimportant since the infrared contamination 
renders even the 
LO result untrustworthy. 

However, the part of the solution governing the 
evolution in $k^2$ is not only infrared safe but is influenced strongly 
by diffusion into the ultraviolet. Hence the effective scale in the
problem 
is greater than $k^2$, and this increase becomes more significant as $x$ 
decreases. This leads to the effective coupling constant decreasing as 
we go to smaller $x$, behaving like $1/(\ln k^2/\Lambda^2)+A(\ln(1/x)/
\ln(k^2/\Lambda^2))^{\half})$ rather than 
$1/(\ln(k^2/\Lambda^2))$. 
This result is quantified by using the BLM scale fixing procedure 
\ref\BLM{S.J. Brodsky, G.P. Lepage and 
P.B. Mackenzie, \PR \vyp{D28}{1983}{228}.} for
both LO and NLO quantities, obtaining precisely the same result of
$A=3.63$ in both cases. It suggests that 
the effective splitting function governing the evolution does not grow like a
power of $x^{-1-\lambda}$ as $x \to 0$, 
but is softened to something of the form 
$(1/x)\exp((\ln(1/x))^{\half})\rho(k^2))$, 
though it seems difficult to obtain the
precise form. This result means that the NLO corrections to the kernel not
concerned with the running of the coupling are also relatively
unimportant 
for the term governing the evolution, simply because the coupling constant
associated with them is so small. Therefore, it seems as though we have 
good predictive power for the evolution of the gluon at small $x$,
but that it is very different from the LO-BFKL prediction with fixed
$\alpha_s$. Because the behaviour of physical structure functions at
small $x$ is related to the gluon via the convolution of a
$k^2$-dependent cross-section at the top of the gluon ladder 
\ref\kti{S. Catani, M. Ciafaloni 
and F. Hautmann, \PL  \vyp{B242}{1990}{97}; \NP \vyp{B366}{1991}{135};
\PL\vyp{B307}{1993}{147}.}\ref\ktii{J.C. Collins and R.K. Ellis, \NP 
\vyp{B360}{1991}{3}.}, all such
effects are associated with the ultraviolet diffusion. Hence, the
evolution of physical quantities is governed by the same effective
coupling constant, and is completely predictive, being somewhat
different from both the LO-BFKL predictions with fixed $\alpha_s$ and
the fixed order in $\alpha_s(Q^2)$ DGLAP descriptions. 

In this paper I will demonstrate the results discussed above. I will
start with
a brief discussion of the LO BFKL solution with fixed coupling, emphasising
the role played by the infrared and ultraviolet regions of transverse 
momentum.
I will then look at the same equation for running coupling, showing how the 
solutions change. This will facilitate a discussion of the real
importance of the total NLO correction to the BFKL equation. 
Finally, I will examine the implications of my results for physical
quantities and
give a brief outline of phenomenological consequences, showing that my 
results work very well when used to analyse experimental data. 
I note that a
very brief account of this work, which nevertheless contains many of the main
ideas, appears in \ref\durhamwg{J.R. Forshaw, G.P. Salam and R.S. Thorne, 
{\tt hep-ph/9812304}, to be published in proceedings of ``3rd UK Phenomenology
Workshop on HERA Physics'', Durham, September 1998.}.            

\newsec{BFKL Equation for fixed $\alpha_s$.}

The BFKL equation for zero momentum transfer is an integral equation for the 
4-point $k_{T}$-dependent gluon Green's function for forward
scattering in the 
high energy limit, $f(k_1,k_2, \alpha_s/N)$ where $N$ is the Mellin conjugate
variable to energy. Throughout this paper I will consider the canonical 
physical process of deep-inelastic scattering where the bottom leg is 
convoluted with a bare gluon density and the top leg with an off-shell 
hard scattering process. Hence, $k_2$ is taken to be some fixed scale 
$Q_0^2$ typical of soft physics\foot{Strictly speaking, within the leading 
twist collinear factorization framework this lower leg should be on-shell,
so $Q_0^2$ is a regularization scale.},
while $k_1^2=k^2$, i.e. a variable scale typically $\gg Q_0^2$. 
In this case $N$ is the conjugate variable to $x$, i.e. we 
define the moment space structure functions by the Mellin transformation, 
\eqn\melltranssf{{\cal F}(N,Q^2)= \int_0^1\,x^{N-1} F(x,Q^2)dx.}
and the moment
space parton distributions as the Mellin transformation
of a rescaled parton density, i.e. 
\eqn\melltranspd{f(N,Q^2)= \int_0^1\,x^{N}{\rm f}(x,Q^2)dx.}

Using these definitions the BFKL equation becomes
\eqn\bfkli{f(k^2, Q_0^2, \bar\alpha_s/N)=f^0(k^2, Q_0^2)+{\bar
\alpha_s \over N}
\int_{0}^{\infty}{dq^2 \over q^2}K^{0}(q^2,k^2)f(q^2),}
where
\eqn\kzero{K^0(q^2,k^2)f(q^2)= k^2
\biggl( {f(q^2)-f(k^2) \over \mid k^2-q^2\mid}
+{f(k^2) \over (4q^4+k^4)^{\half}}\biggr),}
$f^0(k^2,Q_0^2)$ is the zeroth order input and $\bar \alpha_s =(3/\pi)
\alpha_s$.
As a simple choice I take
\eqn\inputdef{f^0(k^2,Q_0^2) = \delta(k^2-Q_0^2),}
i.e. the incoming gluon has a fixed nonzero virtuality. With this definition 
a moment space gluon structure function can be defined as\foot{In this
paper 
I will ignore the singlet quark distribution. This is purely for
simplicity 
and
does not change any of the conclusions at all. In most expressions the 
replacement of $g_B(N,Q_0^2)$ with $g_B(N,Q_0^2)+\fourninths
\Sigma_B(N,Q_0^2)$,
where $\Sigma_B(N,Q_0^2)$ is the bare singlet quark distribution, is
all that
is required to make them completely correct.}   
\eqn\gluondef{{\cal G}(Q^2,N)=\int_{0}^{Q^2}{dk^2\over k^2} f(N,k^2,Q_0^2)
\times g_B(N,Q_0^2),}
where $g_B(N,Q_0^2)$ is the bare gluon distribution as a function of
the {\it factorization scale} $Q_0^2$.\foot{In making this definition of the 
gluon
distribution we have defined a factorization scheme.}

In order to solve this equation it is convenient to take a further Mellin 
transformation with respect to $k^2$, i.e. define 
\eqn\mellin{\tilde f(\gamma,N)=
\int_{0}^{\infty}d k^2 (k^2)^{-1-\gamma} f(k^2, N).}
This leads to the BFKL equation written in the form 
\eqn\bfklii{\tilde f(\gamma,N)=\tilde f^0(\gamma, Q_0^2)+
(\bar \alpha_s/N)  \chi(\gamma)
\tilde f(\gamma, N), }
where $\tilde f^0(\gamma, Q_0^2)=\exp(-\gamma \ln (Q_0^2))$
and $\chi(\gamma)$ is the characteristic function
\eqn\kergam{\chi(\gamma)=2\psi(1)-\psi(\gamma)-\psi(1-\gamma).}
Hence,
\eqn\soli{\tilde f(\gamma,N)={\tilde f^0(\gamma,N) \over
1-(\bar\alpha_s/N) \chi(\gamma).}}  

For asymptotically small $x$ this can be accurately inverted back to
$x$ and $k^2$ space using the saddle point 
technique to give the celebrated result, 
\eqn\solii{{\rm f}(x,k^2) \propto x^{-\lambda} \biggl(
{k^2\over \bar\alpha_s \ln(1/x)}
\biggr)^{\half} \exp
\biggl({-\ln^2(k^2/Q_0^2)\over 56 \zeta(3) \bar\alpha_s \ln(1/x)}
+ \cdots.\biggr),}
where $\lambda =4\ln2\bar\alpha_s$ and $\cdots$ denotes subleading terms 
as $x\to 0$. Hence, we see that the BFKL equation at LO predicts powerlike 
growth in $x^{-\lambda}$ and in $k^2$, as well as a diffusion in $k^2$. 
One can also be a little more systematic and solve for the coefficient
functions and anomalous dimensions for the gluon, it is easy to generalise
\soli\ to give the double Mellin space expression for the gluon
structure function
\eqn\soliii{\tilde {\cal G}(\gamma,N)={\tilde f^0(\gamma,N)
g_B(N,Q_0^2) \over \gamma
(1-(\bar\alpha_s/N) \chi(\gamma))},}  
and 
\eqn\invmell{{\cal G}(Q^2,N)={1\over 2\pi i}
\int_{\half-i\infty}^{\half+i\infty}
d\gamma \exp(\gamma \ln(Q^2)) \tilde {\cal G}(\gamma,N).}
From \soliii\ we see that there are poles when $1-(\bar\alpha_s/N)
\chi(\gamma)=0$. Defining the rightmost solution of this equation by
\eqn\gammazero{\chi(\gamma^0(\bar\alpha_s/N)){\bar\alpha_s \over N}=1,}
we obtain the leading twist solution for the gluon structure function
\eqn\soliv{{\cal G}(Q^2,N)= {1 \over -(\bar\alpha_s/N)\gamma^0
\chi'(\gamma^0)}\biggl({Q^2\over
Q_0^2}\biggr)^{\gamma^0}g_B(N,Q_0^2).}
Hence, $\gamma^0(\bar\alpha_s/N)$ is the anomalous dimension
governing the $Q^2$ evolution of the gluon \ref\Jar{T. Jaroszewicz, \PL  
\vyp{B116}{1982}{291}.}, and 
${\cal R}(\bar\alpha_s/N)\equiv -(\bar\alpha_s/N\gamma^0
\chi'(\gamma^0))^{-1}$
is a type of coefficient function giving the normalization \kti. 
Each of these may be expanded as power series in $(\bar\alpha_s/N)$, which
then lead to power series in $\bar\alpha_s\ln(1/x)$ in $x$-space.
Both are only convergent for $\bar\alpha_s/N<4\ln2$, each 
developing a branch point showing that in $x$-space they grow like 
$x^{-(1)-\lambda}$. 
Using the saddle point technique one may find the asymptotic 
form of the $x$-space splitting 
function and coefficient function finding that
\eqn\split{P^0(x) \to {\bar \alpha_s\over x} x^{-\lambda} \biggl({1\over 
(56\pi\zeta(3))^{\half}
(\bar\alpha_s \ln(1/x))^{3/2}}\biggr),}
and
\eqn\coeff{R(x) = 4\ln 2 \bar \alpha_s x^{-\lambda} \biggl({1\over 
14\pi\zeta(3)
\bar\alpha_s \ln(1/x)}\biggr)^{\half}.}
Therefore, both the anomalous dimension and the coefficient function 
predict powerlike behaviour for the gluon distribution, although the
true input
for the distribution is really $R(x)$ convoluted with $g_B(x,Q_0^2)$ of
course, 
and this leads to the exact form of $R(x)$ being sensitive to the 
input $f^0(N,Q_0^2)$.\foot{In the language of the factorization theorem this 
translates into $R(x)$ being regularization scheme dependent, e.g. if one 
uses dimensional regularization rather than a off-shell gluon, $R(x)$
has a factor of $(\bar\alpha_s\ln(1/x))^{-3/4}$ rather than $(\bar\alpha_s
\ln(1/x))^{-\half}$.} However, this powerlike 
behaviour does not set in until very
small $x$, as may be seen by examining the terms in the expansion for 
each quantity in powers of $\bar\alpha_s\ln(1/x)$.  

It has long been suspected that the diffusion property of the solution to the 
BFKL equation may have serious consequences when working beyond the strictly 
LO framework \liprun\diff\ref\harriman{J.R. Forshaw, P.N. Harriman and
P.N. Sutton, \NP \vyp{B416}{1994}{739}.}\haakman\mueller. 
One may appreciate this by recognising 
that in the small $x$ limit, defining $\xi=\ln(1/x)$, we may write,
\eqn\diffeq{{\rm f}(k^2,Q_0^2,\xi) =\int dr^2 {\rm f}(k^2,r^2,\xi')
{\rm f}(r^2,Q_0^2,\xi-\xi').}
For a given $\xi'$ we can ask for the mean and the deviation of $\ln(r^2)$.
This is equivalent to asking for the typical $\ln(k^2)$ at some point along
the ladder diagram representing the function ${\rm f}(k^2,Q_0^2,\xi)$, 
and also its 
spread, i.e. the range of important values of $q^2$ involved in finding
the solution of the BFKL equation. The result is well known:
\eqn\diffi{<\ln(r^2/(kQ_0))>={\ln(k^2/Q_0^2)\over 2}
\biggl(1-2{\xi'\over \xi}\biggr),}
and the RMS deviation is 
\eqn\diffii{\sigma^2=28\zeta(3)\bar\alpha_s \xi'\biggl(1-{\xi'\over \xi}
\biggr).}
So over much of the ladder $<\ln(r^2)>\sim \half \ln(k^2/Q_0^2)$ and 
$\sigma\sim (14\zeta(3)\bar\alpha_s\ln(1/x))^{\half}$. Hence, for very low
$x$ there will be significant diffusion into both the infrared and the 
ultraviolet. In the case of fixed coupling this does not cause any serious 
problems. However, in the case of the running coupling the size of the 
coupling grows quickly in the infrared, and hence this diffusion suggests 
that there will be serious contamination from nonperturbative physics.
 
Before looking at the BFKL equation for running coupling let us
briefly 
examine
the role played by the various regions of $q^2$ in the fixed coupling case. 
In order to determine the role played by the region of low transverse 
momentum 
we consider a upper cut-off of $k_0^2$ in the integral in \bfkli. The only 
restriction we place on $k_0^2$ is that $k_0^2\ll k^2$ for whatever 
transverse
momentum we ultimately 
wish to consider at the top of the ladder. With this restriction 
we see that for all momenta over which we integrate we have the relation   
\eqn\kzerocut{K^0(q^2,k^2)f(q^2)= f(q^2) 
+{\cal O}\biggl({k_0^2 \over k^2}\biggr),}
and inserting into the cut--off version of \bfkli\ we obtain
\eqn\bfklicut{{\bar \alpha_s \over N} \int_{0}^{k_0^2}{dq^2 \over q^2}
K^{0}(q^2,k^2)f(q^2)=
{\bar \alpha_s \over N} h(k_0^2, f(k^2<k_0^2)) +{\cal O}\biggl({k_0^2 \over
k^2}\biggr).}
The integral over the region $q^2\leq k_0^2\ll k^2$ contributes only a
constant 
to the right hand side of \bfkli, dependent on the form of $f(q^2, N)$ at low
momentum, but independent of the value of $k^2$ we consider as long as it is 
large. If $k^2$ is actually smaller than $k_0^2$ then 
$h(k_0^2, f(k^2<k_0^2))$
becomes a much more sensitive function of $k^2$, 
and in the limit $k^2 \to 0$ it is easy to see 
that it becomes equal to the value of the integral in \bfkli\ with no upper
cut--off. Hence, $h(k_0^2, f(k^2<k_0^2))$ has the same 
structure for $k^2\to 0$ as the full integral on the right hand side of 
\bfkli, but tends to a constant function of $k_0^2$ for $k^2\gg k_0^2$.
 
Thus, if we imagine imposing an infrared cut-off on \bfkli\ we can 
simply subtract the result of the integral up to $k_0^2$ 
(now with a different $f(q^2)$ for low $q^2$, in particular the infrared
cut-off renders it infrared finite) from the right hand 
side of \bfkli, obtaining (up to higher twist corrections).
\eqn\bfkliir{f(k^2, Q_0^2, \bar\alpha_s/N)=f^0(k^2, Q_0^2)
-{\bar\alpha_s\over N}h(k^2,k_0^2)+{\bar \alpha_s \over N}
\int_{0}^{\infty}{dq^2 \over q^2}K^{0}(q^2,k^2)f(q^2).}
Taking the Mellin transform of this equation we get
\eqn\bfkliiir{\tilde f(\gamma,N)=\tilde f^0(\gamma,
Q_0^2)-(\bar\alpha_s/ N) \tilde h(\gamma,k_0^2)
+(\bar \alpha_s/N)  \chi(\gamma) \tilde f(\gamma, N),}
where $\tilde h(\gamma,k_0^2)$ is analytic for $\gamma >0$ ($h(k^2,k_0^2)$
tends to a constant at high $k^2$). 
This second term on the right may simply be absorbed into the definition of 
the input and our expression for $\tilde f(\gamma, N)$ is exactly the same as 
in \soli\
up to this transformed input, i.e. 
\eqn\solicut{\tilde f(\gamma,N)={\tilde f^0(\gamma,N)-(\bar\alpha_s/
N)\tilde h(\gamma,k_0^2) 
\over 1-(\bar\alpha_s/N) \chi(\gamma)}.}  
Performing the inverse Mellin transformation,
then for the leading twist solution the pole is in exactly the same place
and we obtain exactly the same $k^2$ dependence as previously, but a 
potentially very different
$N$-dependent normalization. Hence we see that the region of transverse 
momentum $\ll k^2$ contributes very significantly to the overall normalization
of our leading twist solution, but negligibly to the evolution,
essentially because the 
contribution from the infrared region coming from the convolution in the BFKL
equation is the same for all high $k^2$. We also notice that the
other, higher twist poles found in \soli\ are now eliminated by the
presence of $-(\bar\alpha_s/ N)h(\gamma,k_0^2)$.

This above argument is hardly new, and much more detailed analysis can
be found
in \ref\ircutoffi{J.C. Collins and P.V. Landshoff, \PL \vyp{B276}{1992}{196}.}
\ref\ircutoffii{M.F. McDermott, J.R. Forshaw and G.G. Ross, 
\PL \vyp{B349}{1995}{189} \semi M.F. McDermott and J.R. Forshaw, \NP 
\vyp{B484}{1997}{248}.} who consider the Mellin space solution carefully,
showing that the infrared cut-off does indeed change only the residue of the 
rightmost pole in $\gamma$ (and removes all poles in the left half plane). 
It is also noted that infrared 
cut-offs influence only the normalization of the gluon distribution, leaving 
the shape in $x$ as well as $Q^2$ largely unchanged \harriman. 
This is because the 
effect is to change the type of singularity in $N$-space, but not the actual
position, i.e. $N=4\ln2 \bar\alpha_s$. However, this is not usually 
discussed together with the phenomenon of diffusion. 
In the case of fixed coupling
the effect of diffusion is less important than for running coupling for
the obvious reason that the coupling is the same at all scales. Nevertheless,
the above arguments imply that in the case of running coupling diffusion
into the infrared, i.e. strong coupling, should again only influence the 
normalization of the gluon, while diffusion into the ultraviolet, i.e. 
weak coupling, should only influence the evolution in $Q^2$. 
We will now investigate this in more detail. 

\newsec{BFKL Equation for Running Coupling.}

It was expected in \glr\ that the way to incorporate the running coupling into
the BFKL equation was to modify \bfkli\ to 
\eqn\bfklruni{f(k^2, Q_0^2, \bar\alpha_s(k^2)/N)=f^0(k^2, Q_0^2)+
{\bar \alpha_s(k^2) \over N}
\int_{0}^{\infty}{dq^2 \over q^2}K^{0}(q^2,k^2)f(q^2),}
where
\eqn\coupdef{\alpha_s=1/(\beta_0\ln(k^2/\Lambda^2)),}
$\beta_0 = (11-2N_f/3)/(4\pi)$, and $N_f$ is the number of active 
flavours. 
One of the main results of the NLO corrections to the BFKL kernel is to 
show that this is indeed an effective way to account for the running coupling
(this will be discussed more later). One can solve this equation in the 
same type of way as for the fixed coupling case, i.e. take the Mellin 
transformation with respect to $(k^2/\Lambda^2)$. 
It is most convenient to first multiply through by 
$\ln(k^2/\Lambda^2)$, and then obtain 
\eqn\bfklrunii{{d\tilde f(\gamma,N)\over d \gamma}={d \tilde 
f^0(\gamma, Q_0^2) \over
d\gamma}-{1\over \bar\beta_0 N} \chi(\gamma)
\tilde f(\gamma, N),}
where $\bar \beta_0=(\pi\beta_0/3)$. 
The inclusion of the running coupling has thus completely changed the form of 
our double Mellin space equation, turning it from a simple equality into a 
first order differential equation. However, this may be easily solved to 
give,
\eqn\solruni{\tilde f(\gamma,N)=\exp(-X(\gamma,N)/(\bar\beta_0 N))
\int_{\gamma}^{\infty}
{d \tilde f_0(\tilde \gamma,N,Q_0^2)
\over d\tilde \gamma}\exp(X(\tilde \gamma)/(\bar\beta_0 N))d\tilde\gamma,}
where
\eqn\solrunii{X(\gamma)=
\int_{\half}^{\gamma}\chi(\hat\gamma)d\hat\gamma
\equiv \biggl(2\psi(1)(\gamma-\half)-
\ln\biggl({\Gamma(\gamma) \over \Gamma(1-\gamma)}\biggr)\biggr).}
The leading singularity in the $\gamma$ plane for
$\exp(-X(\gamma)/(\bar\beta_0 N))$, is 
cancelled by an integral from $0 \to \gamma$ of the integrand depending
on $\tilde \gamma$ \janpcoll, and so up to higher twist corrections we may 
simplify \solruni\ to 
\eqn\solruniii{\tilde f(\gamma,N)=\exp(-X(\gamma)/
(\bar\beta_0 N))\int_{0}^{\infty}
{d\tilde f_0(\tilde\gamma,N,Q_0^2)
\over d\tilde \gamma}\exp(X(\tilde \gamma)/(\bar\beta_0 N))d\tilde\gamma.}
Using our previous choice of input, i.e. fixed virtuality, we obtain the 
moment space gluon structure function
\eqn\solruniv{\eqalign{{\cal G}(Q^2,N)&={1\over 2\pi i}
\int_{\half -i\infty}^{\half+i\infty}
{1\over \gamma}
\exp(\gamma\ln(Q^2/\Lambda^2)-X(\gamma)/(\bar\beta_0 N))
d\gamma
\cr
&\hskip 1in \times\int_{0}^{\infty}
\exp(-\tilde\gamma\ln(Q_0^2/\Lambda^2)+X(\tilde \gamma)/
(\bar\beta_0N))d\tilde\gamma
\,g_B(Q_0^2,N)\cr
&=a(Q^2,N)b(Q_0^2,N)g_B(Q_0^2,N).\cr}}

Hence, as in the case of fixed coupling constant, at leading twist 
the solution
has factorized into a $Q^2$-dependent part $a(Q^2,N)$ which determines 
the evolution, and an input dependent part $b(Q_0^2,N)$ 
which can be combined with the bare input gluon 
distribution to provide the input for the gluon distribution 
\jan\janpcoll\ciafrun. This time the 
different parts are not so easy to calculate though.  Clearly the
behaviour of both functions is determined by the form of 
$\exp(X(\gamma)/(\bar\beta_0 N))$,
since this determines the singularity structure. 

Considering $b(Q_0^2,N)$ we find that $\exp(X(\gamma)/(\bar\beta_0
N))$ has poles at all positive 
integers, and zeroes at $0$ and all negative integers. Hence, $b(Q_0^2,N)$ is
not properly defined, since the integrand has an infinite number of poles 
lying along the line of integration. These are due to the divergence of the 
coupling at low $k^2$ and can only be removed by some 
infrared regularization. Hence, the diffusion into the infrared has destroyed
the apparent (limited) predictive power for the input.
Imposing some regularization scale $k_0^2$
and repeating the same arguments as the previous section it is clear that 
up to higher twist corrections the effect of the regularization is simply
to leave the factor $a(Q^2,N)$ unchanged, and change $b(Q_0^2,N)$
to 
\eqn\regin{c(Q_0^2,k_0^2,N)=\int_{0}^{\infty}\biggl({d \tilde 
f_0(\tilde\gamma,N,Q_0^2) \over d \tilde \gamma}+
\tilde h(\tilde\gamma,k_0^2)\biggr)\exp(X(\tilde
\gamma)/(\bar\beta_0 N))d\tilde\gamma,}
where the factor $\biggl({d \tilde f_0(\tilde\gamma 
Q_0^2,N) \over d \tilde \gamma}+\tilde h(\tilde\gamma,k_0^2)\biggr)$ 
removes the singularities in  
$\exp(X(\tilde\gamma)/(\bar\beta_0 N))$. Thus, we have
\eqn\solrunv{{\cal G}(Q^2,N)=a(Q^2,N)c(Q_0^2,k_0^2,N)g_B(Q_0^2,k_0^2,N),} 
as a well-defined solution.
\foot{That the solution at leading twist is of this 
general form was shown in \ciafrun\ by putting the BFKL equation  with
running coupling in the form of an infinite order differential equation with
effective potential depending on the low $k^2$ 
regularization of the coupling.}
For a given regularization one can solve for $c(Q_0^2,k^2_0,N)$, as has 
been done numerically \foot{The numerical solutions are always for the
whole of the gluon structure function, not just $c(Q_0^2,k_0^2,N)$.}, 
generally obtaining some powerlike growth in $x$-space,
but which is totally dependent on the type and scale of regularization 
\hancock\nik\haakman\ref\harrjeff{J.R. Forshaw and P.N. Harriman,
\PR \vyp{D46}{1992}{3778}.}. No
real predictive power remains (this will be discussed 
more in \S 5).

Even without regularization there is no obstruction to solving for the
$Q^2$ dependent part of the gluon distribution, and this is unchanged by this 
regularization, i.e. is unaffected, up to ${\cal O} (k_0^2/Q^2)$
corrections, by the diffusion into the infrared. The 
function $a(Q^2,N)$ is, of course, determined by the singularities of 
$\exp(-X(\gamma)/(\bar\beta_0 N))$ in the $\gamma$ plane.
Here we notice a fundamental difference between the cases of the fixed and 
running couplings. Whereas previously the leading singularity was a pole 
at $(\bar\alpha_s/N)\chi(\gamma)=1$, i.e. at $\gamma\to \half$ as $N \to 
4\ln 2\bar\alpha_s$, now the leading singularity is an essential singularity at
$\gamma=0$: there is no powerlike behaviour in $Q^2$. Similarly, the 
branch point in the $N$ plane at $4\ln 2\bar\alpha_s$ has become an essential
singularity at $N=0$: there is no powerlike behavior in $x$. The introduction
of the running of the coupling has therefore also had an extreme
effect upon the 
evolution, changing its character completely. This point has been noticed
before \janpcoll\haakman, but not emphasised or 
studied in detail. Hence I stress 
the fundamental results of introducing a running coupling: the 
$Q^2$-independent part of the solution is formally divergent, and hence is 
totally regularization scheme dependent: the $Q^2$-dependent part has no
powerlike growth in $x$. 

In fact we can obtain some information about the $x$ behaviour by noting that 
we can find the inverse Laplace transformation of
$\exp(-X(\gamma)/(\bar \beta_0 N))$ precisely \haakman\levin. It is a standard
result that 
\eqn\bess{{1\over 2\pi i}\int_{-i\infty}^{i\infty}\exp(N\xi+K/N)dN=
(A/\xi)^{\half}I_1(2(A\xi)^{\half}),}
were $I_1(z)$ is the modified Bessel function, which for large values of its
argument $\to \exp(z)/(2\pi z)^{\half}$. Hence for large $\xi$
\eqn\asymp{{\rm a}(\xi,\gamma) \sim (-X(\gamma)/\bar \beta_0 \xi)^{3/4}
\exp(2(-\xi X(\gamma)/\bar \beta_0))^{\half}.}
It is difficult to perform the inverse Mellin transformation to get the $Q^2$
dependence, but the leading singularity is at $\gamma=0$. Thus, for any $Q^2$ 
the leading twist solution for ${\rm a}(\xi,Q^2)$ must have small $x$ behaviour
going like $\exp(\xi^{\half})$ rather than the $\exp(\lambda\xi)$ for the fixed
coupling case. This is easy to understand in terms of the diffusion
picture. Since the function ${\rm a}(\xi,Q^2)$ is insensitive to the diffusion 
towards the infrared, but sensitive to that into the ultraviolet, we expect
the typical scale in the process to be determined by this latter diffusion.
Thus the typical scale for the process will be approximately set by 
$\ln (\tilde k^2)\sim \ln(k^2)+\sigma \sim \ln(k^2) +4
(\bar\alpha_s(k^2)\xi)^{\half}$. Hence, the effective strength of the running
coupling will be set by $\tilde k^2$, rather than $k^2$, and as $x\to 0$
we will have $\bar \alpha^{eff}_s\sim 1/(\xi)^{\half}$. This type of effective
coupling has precisely the effect of turning the low $x$ behaviour of
the fixed coupling solution to that which we find for the running
coupling. Hence, the diffusion into the ultraviolet has 
a major impact on the $Q^2$ dependent part of our gluon distribution,
but in a well controlled, and in principle calculable way, unlike the 
effect of the infrared diffusion on the $Q_0^2$ dependent input.

Of course, this is just a qualitative argument giving only the general form of
the results. It is also for the function ${\rm a}(x,Q^2)$, which must be
convoluted with an unknown, if $Q^2$-independent input function in order to 
obtain physical results. It would be nice to be more quantitative, and also
to calculate some physical quantity unambiguously. For example, staying
in moment space we can examine $(d {\cal G}(N,Q^2)/d\ln(Q^2))$, which is 
an entirely perturbatively calculable quantity, and its
transformation into $x$-space. This will be considered in the next section.

\newsec{Solving the BFKL Solution for Running Coupling - Evolution.}

The usual approaches taken to finding the solution for $a(Q^2,N)$ (or the 
full solution) are to 
assume that for small $x$ one can expand $X(\gamma)$ about $\gamma=\half$
to some finite order in $\gamma$,\foot{This is equivalent to writing the 
$k^2$-space BFKL equation as an infinite order differential equation 
and truncating
at a low order in derivatives, or iterrating the LO solution in the
truncated form \soliii\ in the NLO equation.}  usually to 
${\cal O}(\gamma^3)$, or to use the 
saddle point method. Neither of these are at all accurate unless $Q^2$ is very
large indeed. This is because along a line parallel to the imaginary axis
$X(\gamma)$ is not at all well represented by
the first few terms in a power series in $\gamma$ about either $\gamma
= \half$ or about the saddle point. The former can be seen  
in \fig\cubed{Comparison of the full function $\exp(\gamma
\ln(Q^2/\Lambda^2)-X(\gamma)/(\bar\beta_0 N))$ to 
the case where the exponent is truncated 
at ${\cal O}(\gamma^3)$ 
along the line $\Re(\gamma)=\half$. I choose $\ln(Q^2/\Lambda^2)=6$ and
$1/(\bar\beta_0 N) =2$.}, where we compare the the  full function $\exp(\gamma
\ln(Q^2/\Lambda^2)-X(\gamma)/(\bar\beta_0 N))$ to 
the case where the exponent is truncated 
at ${\cal O}(\gamma^3)$ 
along the line $\Re(\gamma)=\half$.\foot{Actually I 
plot the real part of the functions. The imaginary part is odd and integrates
to zero.} Clearly the 
integral over the two contours need bear little similarity. 

When using the saddle point technique one finds the minimum of the exponent 
of the integrand in the definition of $a(Q^2, N)$ and expands in a 
Taylor series about this point. This minimum occurs when 
\eqn\min{{d\over
d\gamma}(\gamma\ln(Q^2/\Lambda^2)-X(\gamma)/(\bar\beta_0 N))=0, }
which using the definition \solrunii\ leads to 
\eqn\minii{{1\over 
\bar\beta_0\ln(Q^2/\Lambda^2)N}\chi(\bar\gamma)\equiv{\bar\alpha_s(Q^2)\over
N}\chi(\bar\gamma)=1,}
i.e. at $\bar\gamma=\gamma^0(\bar\alpha_s(Q^2)/N)$, the anomalous dimension 
for the fixed coupling case, but with the running coupling evaluated
at scale $Q^2$.
The integrand defining $a(Q^2,N)$ is thus evaluated along the axis 
$\Re(\gamma)=\gamma^0(\bar\alpha_s(Q^2)/N))$, i.e.
\eqn\sadi{\eqalign{a(Q^2,N)={1\over 2\pi i}\exp\biggl(\int^{Q^2}&\gamma^0(\bar
\alpha_s(q^2)/N) d\ln q^2\biggr)\cr
&\int_{-i\infty}^{i\infty}
{1\over \gamma^0+\gamma}
\exp(\gamma\ln(Q^2/\Lambda^2) +(X(\gamma_0)-X(\gamma^0+\gamma))
/(\bar \beta_0 N))d\gamma.\cr}}
Letting, $\gamma\to -i\gamma$ and expanding about $\gamma^0(\bar\alpha_s/N)$
this becomes 
\eqn\sadii{a(Q^2,N)={1\over 2\pi }\exp\biggl(\int^{Q^2}\gamma^0(\bar
\alpha_s(q^2)/N)\biggr) d\ln q^2\int_{-\infty}^{\infty}
\biggl({1\over \gamma^0}+\cdots
\biggr)
\exp(\gamma^2\chi'(\gamma^0)/(2\bar\beta_0 N)+\cdots)d\gamma.}
This is then normally evaluated by ignoring all those parts not explicitly
included above, and performing the gaussian integral 
\ref\rsadpoint{A.D. Martin and J. Kwiecinski, \PL \vyp{B353}{1995}{123}.}
obtaining\foot{A
factor of $(\bar\beta_0 N/2\pi)^{\half}$ is absorbed into
$b(Q_0^2,N)$.} 
\eqn\sadiii{a^{SP}(Q^2,N)= {1 \over \gamma^0(\bar\alpha_s(Q^2)/N)(-\chi'(
\gamma^0(\bar\alpha_s(Q^2)/N)))^{\half}}\exp\biggl(\int^{Q^2}\gamma^0(\bar
\alpha_s(q^2)/N) d\ln q^2\biggr).}
This is of the same form as \soliv, i.e. an evolution term governed by the
previous anomalous dimension and a coefficient function which is a power 
series in $\bar\alpha_s/N$, except that now $\alpha_s$ runs with $Q^2$ 
rather than being fixed. This could be taken to imply that one can simply 
extract anomalous dimensions and coefficient functions from this solution 
and that the appropriate scale to use for the coupling is $Q^2$.

The invalidity of this assumption is related to the fact that \sadiii\ is 
in fact a very poor approximation to the full solution for $a(Q^2,N)$.
This is clear because in $x$-space both the perceived anomalous dimension and 
coefficient function above grow like $x^{(-1)-4\ln2\bar\alpha_s(Q^2)}$ 
as $x \to 0$, whereas we know that the complete solution for $a(Q^2,x)$ 
has no real
powerlike behaviour in $x$. We can see how we have obtained such a poor 
approximation  by using the saddle point technique if we examine the form of
the complete integrand along our contour of integration compared with 
the function we have actually integrated  making the approximation in \sadii.
This is seen in \fig\saddle{Comparison of the full function $\exp(\gamma
\ln(Q^2/\Lambda^2)-X(\gamma)/(\bar\beta_0 N))$ to 
the function appearing in the saddle point estimate  
along the line $\Re(\gamma)=\gamma^0(\bar \alpha_s/N)$. 
I choose $\ln(Q^2/\Lambda^2)=6$ and
$1/(\bar\beta_0 N) =2$, so $\gamma^0(\bar \alpha_s/N)=0.384$.},
\foot{Again I plot only the real part.}
and it is glaringly obvious that the saddle point estimate is not at all 
reliable in this case. 
Formally the corrections ignored in evaluating \sadii\
are of higher order in $\beta_0\alpha_s(Q^2)$ than the terms calculated, but
their coefficients grow quickly, i.e. like factorials, 
and to be precise they are powers of 
$\beta_0\alpha_s(Q^2)(\bar\alpha_s(Q^2)\xi)^{r}$ higher than the presented
results, where $r$ is a positive number, and are thus dominant for low
enough $x$. Hence a resummation is really
necessary for a true understanding.  

However, an alternative view of the result in \sadiii\ may lead us towards the
correct physics. It is not really useful to interpret the prefactor in 
this equation as a coefficient function which tells us something about the
normalization of the gluon structure function since $a(N,Q^2)$ must be
multiplied
by an unknown $N$-dependent function in order 
to obtain this distribution. Rather, it is
better to acknowledge that the only real information contained in $a(N,Q^2)$
is on the evolution of the structure function, i.e.
\eqn\evol{{d \ln{\cal G}(N,Q^2) \over d\ln(Q^2)} =
{d\ln a(N,Q^2) \over d \ln(Q^2)}\equiv \Gamma(N,Q^2).}
Thus, using $a(N,Q^2)$ in \evol\ gives us an entirely perturbative effective
anomalous dimension governing the evolution of the gluon distribution. 
Using \sadiii\ we obtain
\eqn\nloevol{\Gamma(N,Q^2)=\gamma^0(\bar\alpha_s(Q^2)/N)-\beta_0\alpha_s(Q^2)
\biggl({d \gamma^0 \over d\ln(\alpha_s)}\biggl({-\chi''(\gamma^0) \over
2\chi'(\gamma^0)}-{1\over \gamma^0}\biggr)\biggr)+{\cal
O}((\beta_0\alpha_s(Q^2))^2) r(\bar\alpha_s(Q^2)/N).}
So within the framework of the LO BFKL equation with running coupling our
unambiguous effective anomalous dimension is the naive leading order result
with coupling at scale $Q^2$ plus a series of corrections going like powers
of $\beta_0\alpha_s(Q^2)$. 

It is tempting to interpret the whole solution for
$\Gamma(\alpha_s(Q^2),N)$ as simply telling us the appropriate scale
to use in the coupling constant for the normal LO result. Indeed, this is the
philosophy in the BLM scheme \BLM\ for scale fixing which uses the NLO 
$\beta_0$-dependent corrections for any process to determine the scale to 
use for the coupling in the LO expression. However, in this case of an
anomalous dimension for
a structure function we have to decide whether it is appropriate to do this 
in $N$-space or $x$-space, i.e. should we write 
\eqn\nblm{{d {\cal G}(N,Q^2) \over d\ln Q^2} \approx 
\Gamma(N,\bar\alpha_s(s(N)Q^2)) G(N,Q^2),}
or 
\eqn\xblm{{d G(x,Q^2) \over d\ln(Q^2)} \approx \int_x^1
P(z,\bar\alpha_s(s(z),Q^2))G(x/z,Q^2)dz.}
Since the moment space expressions are less physical, being defined only by
analytic continuations over much of the $N$-plane we choose the
latter.\foot{Fixing the scale in $N$-space would lead to a scale which
was singular at $N=\lambda(Q^2)$, which does not seem a sensible
proposition, while in $x$-space it is a smooth function of $x$ as we
will see.} As we will see later, this decision is backed up by higher
order calculations. Note that both $\Gamma(N,Q^2)$ and $P(x,Q^2)$ are
entirely independent of factorization scale, and are functions only of
renormalization scale. Indeed, if there were a direct probe of the gluon, i.e.
$G(x,Q^2)$ were directly measurable, then both  $\Gamma(N,Q^2)$ and $P(x,Q^2)$
would be physically defined quantities. As such the choice of the 
renormalization scale is entirely open.  

The simplest thing we can do is to choose the scale for the coupling
constant in the leading order expression so that the NLO term in the $x$-space
version of \nloevol\ is exactly produced by the expansion about 
$\alpha_s(Q^2)$. Thus, writing this $x$-space expression as 
\eqn\nlosplit{(x/\bar\alpha_s(Q^2))P(x,Q^2)=p^0(\bar\alpha_s(Q^2)\xi)
-\beta_0\alpha_s(Q^2)\hat
p^1(\bar\alpha_s(Q^2)\xi)+{\cal O}((\beta_0\alpha_s(Q^2))^2) 
r(\bar\alpha_s(Q^2)
\xi),}
this is the same as 
\eqn\absorb{(x/\bar\alpha_s(Q^2s(\xi
\bar \alpha_s(Q^2))))P(x,Q^2)=p^0(\bar\alpha_s(Q^2s(\xi
\bar \alpha_s(Q^2)))+{\cal O} 
((\beta_0\alpha_s(Q^2))^2) \hat r(\bar\alpha_s(Q^2)\xi),}
if we choose 
\eqn\scale{\exp(s(\xi\bar\alpha_s(Q^2))) ={\hat p^1(\bar\alpha_s(Q^2)\xi)
\over (d p^0(\bar\alpha_s(Q^2)\xi) /d\ln\alpha_s(Q^2))}.}
This is the usual BLM scale fixing, but here we have extra information since,
in principle at least, we know higher order terms and we would expect 
$\hat r(\bar\alpha_s(Q^2)\xi)$ to be small if the scale fixing is correct.

Equation \scale\ can be solved for arbitrary $x$, but it is first useful to 
examine the limit of $x \to 0$ in order to see if our previous expectations
based on qualitative arguments are confirmed. 
Hence we need each of the terms in 
\scale\ in this limit. As $x \to 0$
\eqn\losplit{p^0(\bar\alpha_s(Q^2)\xi) \to {1\over 
(56\pi\zeta(3))^{\half}} \exp(\lambda(Q^2)\xi)(\bar\alpha_s(Q^2)\xi)^{-3/2},}
and therefore
\eqn\losplitderiv{{d p^0(\bar\alpha_s(Q^2)\xi) \over d\ln(\alpha_s(Q^2))}
\to {4\ln 2 \over (56\pi\zeta(3))^{\half}}\exp(\lambda\xi)
(\bar\alpha_s(Q^2)\xi)^{-\half}.}
In order to calculate the $x\to 0$ limit of $\hat p^1(\bar\alpha_s(Q^2)\xi)$
it is easiest to first consider its moment space analogue, i.e. the second
term on the right of \nloevol. First we note that using \gammazero\ 
\eqn\reli{ {d \gamma^0(\bar\alpha_s(Q^2)/N) \over d \ln(\alpha_s(Q^2))}
=-{\chi(\gamma^0(\bar\alpha_s(Q^2)/N)) \over \chi'(\gamma^0(\bar\alpha_s(Q^2)
/N))}.}
In the limit $x \to 0$, $\chi(\gamma^0) \to 4\ln2$ and $\gamma^0\to \half$,
but to be more precise,
\eqn\relii{ \chi(\gamma^0) \to 4\ln2 -14\zeta(3)(\half-\gamma^0)^2 +\cdots.}
Therefore,
\eqn\reliii{\chi'(\gamma^0) \to -28\zeta(3)(\half-\gamma^0)\equiv
-28\zeta(3)\delta\gamma^0.}
Hence,
\eqn\reliv{{d \gamma^0 \over d\ln(\alpha_s)}=-{\chi(\gamma^0) \over
\chi'(\gamma^0)}\to {\ln 2 \over 7\zeta(3)\delta\gamma^0}.}
Since $\delta\gamma^0$ is vanishingly small as $x\to 0$ 
we see that the $1/\gamma^0$ term
in \nloevol\ becomes subleading to the $\chi(\gamma^0)/\chi'(\gamma^0)$ term.
The $N$-space version of $\hat p^1(\bar\alpha_s(Q^2)\xi)$ is thus
$(\chi(\gamma^0)\chi''(\gamma^0)/2(\chi'(\gamma^0)^2)$. To progress further
we need $\delta\gamma^0$ as a function of $N$. This can be obtained by
solving \gammazero\ using \relii. This gives
\eqn\relv{\delta\gamma^0(\bar\alpha_s(Q^2)/N)=\biggl({2\ln 2 \over 7\zeta(3)}
\biggr)^{\half}\biggl({N \over \lambda(Q^2)} -1\biggr)^{\half}.}
This can be substituted into the moment space analogue of $\hat p^1(\bar
\alpha_s(Q^2)\xi)$ and the inverse transformation performed to give
\eqn\nlosplit{\hat p^1(\bar\alpha_s(Q^2)\xi) \to \ln2 
\exp(\lambda(Q^2)\xi).}
This now makes it trivial to solve \scale, and we find that in the coupling in
our LO splitting function
\eqn\expscale{\ln(Q^2/\Lambda^2)\to \ln (\tilde Q^2/\Lambda^2)
=\ln(Q^2/\Lambda^2) +
{(56\zeta(3)\pi)^{\half}\over 4}(\bar\alpha_s(Q^2)\xi)^{\half}.}
This is exactly the sort of scale change we would expect from the diffusion
into the ultraviolet. It also leads to $xP(\bar\alpha_s(Q^2),x)\sim
\exp(1.14(\xi/\bar\alpha_s(Q^2))^{\half})$ as $x\to 0$, precisely
the sort 
of behaviour we would expect from the qualitative discussions in the last 
section. 

We can also solve \scale\ exactly rather than relying on 
asymptotic limits using the power series expansions of 
${d p^0(\bar\alpha_s(Q^2)\xi) \over d\ln(\alpha_s(Q^2))}$ and 
$\hat p^1(\bar\alpha_s(Q^2)\xi)$ in $\bar\alpha_s(Q^2)\xi$. 
The results of such solutions are shown in 
\fig\gluscale{The effective coupling constant for $N_F=4$ 
for the gluon structure
function as a function of $x$ compared to the constant values at the
relevant values of $\ln(Q^2/\Lambda^2)$.},
where I plot the effective coupling constant for $N_f=4$ 
derived as a function of
$x$ compared to its constant value taking $Q^2$ as the scale. The
qualitative result is entirely consistent with \expscale\ though the
effective scale is a little smaller than this asymptotic result at
highish $x$ values due to $p^0(\alpha_s(Q^2)\xi)$ and $\hat
p^1(\alpha_s(Q^2)\xi)$ not yet having reached their asymptotic expressions.  

Hence, this BLM scale fixing procedure leads to a choice of scale which fits 
in well with our naive expectations, and must be at least broadly 
qualitatively correct since it does destroy the powerlike behaviour we get 
from fixed order calculations in $\alpha_s(Q^2)$. Ignoring for the moment 
the fact that we have assumed the manner in which to take account of running
coupling effects in the BFKL equation (we will discuss possible
corrections later), we would still
like to know whether our prescription is a true representation of the full
effect of the running coupling, i.e. whether $\hat r(\bar\alpha_s(Q^2)\xi)$ in 
\absorb\ is really small. At each order in $\beta_0\alpha_s(Q^2)$ it is 
possible to calculate the leading behaviour in the limit $x\to 0$. By power
counting one can see that these leading terms come from keeping only the next 
term not explicitly shown in the exponential in \sadii, i.e. the leading
behaviour is given by        
\eqn\sadiv{a(Q^2,N)={1\over 2\pi \gamma^0}\exp\biggl(\int^{Q^2}\gamma^0(\bar
\alpha_s(q^2)/N)\biggr) d\ln q^2\int_{-\infty}^{\infty}
\exp((\bar\beta_0N)^{-1}(\half\gamma^2\chi'(\gamma^0)+(i/3)\gamma^3
\chi''(\gamma^0)))d\gamma.}
Expanding the exponential in $i\gamma^3$ and performing each relevant integral
gives the most singular behaviour in $(N-\lambda(Q^2))$
at each order in $\beta_0\alpha_s(Q^2)$. This is a series of the form
\eqn\expexpan{a^{SP}(Q^2,N)\sum_{n=0}^{\infty}(-1)^n L_n 
(\beta_0\alpha_s(Q^2))^n (\lambda(Q^2)/(N-\lambda(Q^2)))^{3n/2},}
where asymptotically $L_n\sim (1.84)^nn!$.
Inserting this into \evol\ and performing the transformation into 
$x$-space leads to a power series of the form
\eqn\expexpand{p^0(x,\bar\alpha_s(Q^2))\sum_{n=0}^{\infty}(-1)^n 
A_n (\beta_0\alpha_s(Q^2))^n(\lambda(Q^2)\alpha^{\half}_s(Q^2)\xi^{3/2})^n,}
where the $A_n$ are all positive. If $A_n$ were equal to $3.63^n/n!$ the above 
series would simply be $\exp(-(\lambda(Q^2) 3.63\beta_0\alpha^{3/2}_s(Q^2)
\xi^{3/2})$,
which would be precisely the leading correction in the exponent of 
$p^0(x,\bar\alpha_s(Q^2))$ expected using my choice of scale, i.e.
\eqn\expexpandi{\exp(\lambda(Q^2)\xi)\to \exp(\lambda(Q^2)\xi-
\lambda(Q^2)\xi(3.63\beta_0\alpha^{3/2}_s(Q^2)\xi^{\half})+\cdots).}
In practice this works reasonably well. $A_1=3.63$ of course, since this set
our scale. $A_2=7.08$ rather than $6.59$, and the terms then slowly 
increase above $(3.63)^n/n!$. As $n\to \infty$, $A_{n+1}/A_n 
\to 1.67/n^{\half}$, and therefore \expexpand\ cannot be precisely of the 
suggested form. Nevertheless, it defines a convergent series in 
$(\beta_0\alpha_s(Q^2))^n(\lambda(Q^2)\bar\alpha^{\half}_s(Q^2)\xi^{3/2})$ 
which for a
wide range of values mimics the desired exponential
$\exp(-\lambda(Q^2)\xi(3.63\beta_0\alpha^{3/2}_s(Q^2)\xi^{\half})$ well.  

This above check is not really terribly useful since the right hand
side of \expexpandi\ hardly matches $\exp(\lambda(\tilde Q^2)\xi)$
well for very large $\xi$, and many other terms are important at all
$\xi$. Including our scale choice in the LO 
expression for the splitting function also leads to terms not explicitly shown
in \expexpandi\ (and in the expansion of the unexponentiated terms in 
$p^0(x,\bar\alpha_s(Q^2))$) which are subleading in $\xi$ at each power of
$\beta_0\alpha_s(Q^2)$ to those discussed above. There are also terms of 
this type generated by the subasymptotic corrections to \expscale. In 
principle one could compare with terms generated form a more careful solution 
of \sadii, including also the nonleading parts coming from
\sadiv. This 
rapidly becomes extremely complicated indeed. It appears as though the 
logarithm of the 
splitting function is indeed an oscillating power series in $\beta_0
\alpha_s(Q^2)(\alpha_s(Q^2)\xi)^{\half}$, but it is difficult to prove this 
rigorously. (We do know that the series will converge, or at least be 
unambiguously summable, since the integral defining $a(Q^2,N)$ is well
defined.) The best check to be done at the moment is to calculate the 
whole of the ${\cal O}((\beta_0\alpha_s(Q^2))^2)$ contribution to the 
splitting function exactly, and compare this to that expected if the scale 
choice is correct, i.e.
\eqn\nloscale{\half (\beta_0\alpha_s(Q^2))^2
\biggl({\partial^2 p^0(x,\bar\alpha_s(Q^2))\over \partial (\ln 
\alpha_s(Q^2))^2} + 2{\partial p^0(x,\bar\alpha_s(Q^2))\over \partial (\ln 
\alpha_s(Q^2))}\biggr)
\biggl({\hat p^1(x,\bar\alpha_s(Q^2))\over \partial p^0(x,\bar\alpha_s(Q^2))/
 \partial (\ln \alpha_s(Q^2))}\biggr)^2.}
The relevant terms in $a(N,Q^2)$ can be found by 
considering the terms in \sadii\ multiplying the gaussian which go like 
$\gamma^6/(\bar \beta_0N)^2$ and $\gamma^4/(\bar\beta_0N)$, 
performing the gaussian 
integrals and using the equality $N=\bar\alpha_s(Q^2)\gamma^0$. This gives
\eqn\sadv{a(Q^2,N)=a^{SP}(Q^2,N)\biggl(1-\beta_0\alpha_s(Q^2)\biggl({5 
(\chi''(\gamma^0))^2\chi(\gamma^0)\over 24 (-\chi'(\gamma^0))^3}-
\bigl({\chi''(\gamma^0) \over 2 \gamma^0} -{\chi'''(\gamma^0) \over 8}\bigr)
{\chi(\gamma^0) \over (-\chi'(\gamma^0))^2}\biggr)+\cdots\biggr).}
Inserting into \evol\ and making the transformation to $x$-space we obtain 
the required ${\cal O}((\beta_0\alpha_s(Q^2))^2)$ splitting function
$\hat p^2(\bar\alpha_s(Q^2)\xi)$. This is 
compared to \nloscale\ in \fig\nnlocomp{Comparison of the exact NNLO
splitting function $\hat p^2(\bar\alpha_s(Q^2)\xi)$ with the value
predicted from the choice of scale in the LO expression. Both terms
are weighted by $(\exp(\lambda(Q^2)\xi))^{-1}$.}, where each term is 
weighted by 
$(\exp(\lambda(Q^2)\xi))^{-1}$, and the upper limit of
$\bar\alpha_s(Q^2)\xi
=4$ is chosen since the first $20$ terms in the series expansions of
each expression give a very
accurate representation up to this value and it easily covers the range 
relevant for comparison to HERA data. As one can see, above 
$\bar\alpha_s(Q^2)\xi=1$ the ansatz for the ${\cal
O}((\beta_0\alpha_s(Q^2))^2)$
contribution of the splitting function matches extremely well to the 
explicitly calculated value. Below $\bar\alpha_s(Q^2)\xi=1$ the
matching is not so good, but this is relatively unimportant since in
this region this contribution to the total splitting function is small
compared to the more leading contributions, i.e. the scale change is
quite small and $p^0(\bar\alpha_s(Q^2)\xi)\gg
\beta_0\alpha_s(Q^2) \hat p^1(\bar\alpha_s(Q^2)\xi)\gg
(\beta_0\alpha_s(Q^2))^2 
\hat p^2(\bar\alpha_s(Q^2)\xi)$. In this region the scale choice 
is also sensitive to the interference
with the finite $x$ effects at fixed order in $\alpha_s(Q^2)$ 
which are ignored using this expansion
scheme. Hence, it seems reasonable to conclude that explicit checks
strongly support the assumption that all the running coupling effects
in the evolution can be accurately described by the use of the
effective scale obtained by solving \scale\ in the LO effective
splitting function.

\newsec{Solving the BFKL Equation for Running Coupling - Input.}

We could also attempt to evaluate $b(Q_0^2,N)$ in the same manner, i.e. 
expanding about $\gamma^0(\bar\alpha_s(Q_0^2)/N)$ and calculating an order by 
order series in $\beta_0\alpha_s(Q_0^2)$. 
Of course, without an infrared regulator
we know that $b(Q_0^2,N)$ must be divergent because the integrand has
singularities along the contour of integration, i.e. at integer values of 
$\tilde \gamma$, which lead to ambiguities of order
$(\Lambda^2/Q_0^2)^n$,
i.e. higher twist. 
These singularities do indeed show up in this power series solution.
Expanding about $\gamma^0(\alpha_s(Q_0^2)/N)$ and only keeping the 
lowest order terms one obtains a sensible solution, i.e.           
\eqn\sadqo{b^{SP}(Q_0^2,N)= {1 \over (\bar\alpha_s(Q_0^2)/N)(-\chi'(
\gamma^0(\bar\alpha_s(Q_0^2)/N)))^{\half}}\exp\biggl(\int^{-Q_0^2}\gamma^0
(\bar\alpha_s(q^2)/N) d\ln q^2\biggr).}
Going beyond this approximation one obtains the same sort of series as
for $a(N,Q^2)$, except
that because the contour is now along the real axis, rather than parallel to
the imaginary axis, the terms in the series are all of the same sign rather
than oscillating. This leads to at least one power series behaviour of
the form
\eqn\renorm{\sum_{n=0}^{\infty}B_n n !(\beta_0\alpha_s(Q_0^2))^n
(\alpha_s(Q_0^2)/(N-\lambda(Q_0^2)))^{n/2},}
where $B_n$ is roughly $B^n$, and $B\sim 3.6$. One can take
the inverse transformation of this 
series term by term, obtaining a power series in $x$-space which sums
to approximately the form    
\eqn\renormi{\exp(\lambda(Q_0^2)\xi)
\sum_{n=0}^{\infty}(B\beta_0\alpha^{3/2}_s(Q_0^2)\xi^{\half})^n.}
Hence, in this case the power series is suggestive of the fact that due to the 
diffusion into the infrared the appropriate coupling for
$b(Q_0^2,N)$ depends not simply on $\ln(Q_0^2/\Lambda^2)$ but on 
$\ln(Q_0^2/\Lambda^2)-3.63\beta_0(\alpha_s(Q_0^2)\xi)^{\half}$, the exact 
opposite of the case for $a(Q^2,N)$. 

Of course, the infrared diffusion is a 
rather complicated problem, and the series in \renormi\ is only
convergent for $(B\beta_0\alpha^{3/2}_s(Q_0^2)\xi^{\half})\leq 1$. This
indicates that I have been too simplistic in transforming \renorm\ to 
$x$-space term by term. The series in \renorm\ is not defined for any $N$,
and before going to $x$-space we must solve this problem. The series \renorm\ 
may be summed using standard Borel transformation techniques. This leads to 
a well-defined series up to an ambiguity of the form 
$\exp(-(N-\lambda(Q_0^2))^{\half}/
(B\beta_0\alpha^{3/2}_s(Q_0^2))$. Now performing
the transformation to $x$-space we obtain a well-behaved series in       
$(B\beta_0\alpha^{3/2}_s(Q_0^2)\xi^{\half})$ as well as an ambiguity of
order
$\exp(\lambda(Q_0^2)\xi)\exp(-1/(B^2\beta^2_0\alpha^{3}_s(Q_0^2)\xi))$,
where $B^2\approx 13$. This latter ambiguity 
is due to the presence of an infrared renormalon 
\ref\renormalon{V.I. Zakharov, \NP \vyp{B385}{1992}{452}\semi
M. Beneke, \NP \vyp{B405}{1993}{424}.} in the expression for
$b(Q_0^2,N)$, and will be cancelled by similar ambiguities in higher twist 
corrections.\foot {The ambiguity is seen as the nonperturbative contribution 
to the solution in \armesto.}
Such terms are therefore taken to estimate the size of 
higher twist effects. In this case we see that due to diffusion becoming
enhanced at small $x$, this infrared induced uncertainty quickly becomes
large at small $x$, and indeed the calculation of the normalization of the 
gluon Green's function is only at all reliable in the limit
\eqn\renormii{13\beta^2_0\alpha^{3}_s(Q_0^2)\xi \ll 1.}
Hence, we find that even if we had a reliable model for the bare gluon 
distribution $g_B(Q_0^2,N)$\foot{Given that the function $b(Q_0^2,N)$
is dependent on the type of collinear regularization as well as
the ambiguity discussed above this actually seems rather unlikely.} 
we cannot calculate the input for the gluon 
distribution at small $x$ within perturbation theory, and previous 
conclusions on the infrared diffusion physics ruining perturbative 
predictability \mueller\mplusk\ are confirmed. In particular we 
note that the requirement in \renormii\ is basically identical to that found 
in \mueller\mplusk, and indeed, if the series in eq. (45) of the
latter is summed it has an ambiguity of exactly the same type as discussed 
above (though in \mplusk\ the series in $x$-space was found directly).
However,
here I stress that this ambiguity is unique to the normalization function,
and does not affect the evolution, which is calculable in 
perturbation theory.  

Before finishing this section it is interesting to discuss the relationship
between the solutions obtained via the techniques in this paper and solutions
obtained by other authors. It has been noted by several authors 
(\mplusk\levin\armesto) that the asymptotic solution for the BFKL equation 
with running coupling has the general form,
\eqn\BFKLapp{{\rm f}(Q^2,Q_0^2,\xi)\sim
{1\over (\alpha_s\xi)^{\half}}\exp\biggl(
\lambda(QQ_0)\xi +K\beta_0^2\alpha_s^5\xi^3-{\ln^2(Q^2/Q_0^2) \over
56\zeta(3)\bar\alpha_s\xi}+\cdots\biggr),}
where unless explicitly stated $\alpha_s$ is at some fixed scale $\mu$,
and $K=(7/6)\zeta(3)(3/\pi)^3(4\ln2)^2$.
This seems rather at odds with the results discussed above. However, it is 
not difficult to see how this solution appears. Ignoring the term in the
exponent going like $\alpha^5_s\xi^3$ one achieves a solution of this form 
simply by taking the transformation to $x$-space of the product 
$a^{SP}(Q^2,N)b^{SP}(Q_0^2,N)$ in the limit $x \to 0$, and only keeping 
the most dominant terms in the series expansions of the couplings about 
scale $\mu$.       

It is not too much extra work to see where the  $\alpha^5_s\xi^3$
terms come from.
Consider if rather than taking the saddle point approximations for $a(Q^2,N)$ 
and $b(Q_0^2, N)$ one takes the solution of \sadiv\ for $a(Q^2,N)$ and the 
equivalent expansion for $b(Q_0^2,N)$. The solution for $b(Q_0^2,N)$ in this
approximation is of precisely the same form as \expexpan\ once we replace
$a^{SP}(Q^2,N)$ with $b^{SP}(Q_0^2,N)$ and remove the factors of $(-1)^n$
(the $L_n$ are identical). If we multiply the two series in these
expressions together then 
since at large orders $L_n\sim a^nn!$, the resulting series is to good 
accuracy proportional to
\eqn\expexpani{\sum_{n=0}^{\infty}L_{2n} 
((\beta_0\alpha_s)^2 (\lambda/(N-\lambda))^{3})^n,}
where $\alpha_s\equiv \alpha_s(\mu^2)$, i.e. we expand $\alpha_s(Q^2(Q_0^2))$ 
about $\alpha_s(\mu^2)$, and
asymptotically $L_{2n+2}/L_{2n}\to (63\zeta(3)/(8\ln 2))n^2$. 
Multiplying this by the two saddle point solutions, and performing the 
transformation to $x$-space this sum introduces precisely 
$\exp(K\beta_0^2\alpha_s^5\xi^3)$ with the correct value of $K$. Hence, this
non-Regge term comes about due to interference between the input term 
$b(Q_0^2,N)$ and the evolution term $a(Q^2,N)$. 

Hence, these previous results do appear by taking the transformation to 
$x$-space of the product of truncated solutions for $a(Q^2,N)$ and 
$b(Q_0^2,N)$. However, I 
would argue that these solutions are not representative of any 
real physics, since neither of these truncations is at all accurate
except at quite high $x$. For given $Q_0^2$ \BFKLapp\ is only applicable for 
$x$ satisfying \renormii, in which case the $x\to 0$ approximations
used to derive \BFKLapp\ are generally rather inaccurate. 
When \renormii\ is not satisfied the 
transformation of $b(Q_0^2,N)$ is indeterminate, and that of
$a(Q^2,N)$
requires resummation. The only sensible 
option seems to be to factor out $b(Q_0^2,N)$ and simply use $a(Q^2,N)$ to
determine the evolution as accurately as possible, rather than trying to find 
${\rm f}(Q^2,Q_0^2,\xi)$. Then we know from the general arguments already
discussed that the Regge term $\exp(\lambda\xi)$ is nothing to do with the
true result, let alone the non-Regge term $\exp(K\beta_0^2\alpha_s^5\xi^3)$.  

\newsec{NLO Corrections to the BFKL Equation.}

So far I have simply assumed that an accurate way to account for the running
of the coupling in the LO BFKL equation is to use \bfklruni. This is an 
assumption which involves the resummation of an infinite number of terms, i.e.
it assumes that at all orders in $\alpha_s(\mu^2)$ the dominant contribution
to the BFKL equation due to the running coupling is 
\eqn\runassump{{\bar \alpha_s \over  N}(-1)^n
(\beta_0\alpha_s(\mu^2)\ln(k^2/\mu^2))^n 
\int_0^{\infty}{dq^2\over q^2}K^0(q^2,k^2)f(q^2),}
Until recently this has been an assumption for all $n\geq1$ although
the 
above terms must be present. However, the 
recent calculation of the NLO correction to the BFKL equation has given us 
some insight into this question. Formally the NLO BFKL equation may be
written as   
\eqn\NLOBFKL{\eqalign{f(k^2,Q_0^2,\bar\alpha_s(\mu^2)/N)= f^0(k^2,Q_0^2)
&+\biggl({\bar\alpha_s(\mu^2) \over N}\biggr)
\int_0^{\infty}{dq^2\over q^2}(K^0(q^2,k^2)\cr
&-\beta_0\alpha_s(\mu^2)
\ln(k^2/\mu^2)K^0(q^2,k^2)-\alpha_s(\mu^2)K^1(q^2,k^2))f(q^2),\cr}}
where $K^1(q^2,k^2)$ can be found in \NLOBFKLlf. This is the strictly NLO
equation with no resummation at all. 
The separation of the NLO part into the running coupling part and the part
depending on $K^1(q^2,k^2)$ is arbitrary. The former is the first term 
in the infinite series we have already considered, but the latter also 
contains some pieces which may be associated with the running of the coupling,
i.e. going like $\beta_0$.
  
This equation can be solved using the same 
methods which were applied in \S 4. Taking the Mellin transformation,
this time with respect to $(k^2/\mu^2)$ we obtain
\eqn\nlomell{\tilde f(\gamma,N)=\tilde f^0(\gamma,Q_0^2)+
\biggl({\bar\alpha_s(\mu^2)
\over N}\biggr)\biggl((\chi_0(\gamma)
-\alpha_s(\mu^2)\chi_1(\gamma))\tilde f(\gamma,N)+\beta_0\alpha_s(\mu^2)
{d (\chi_0(\gamma)\tilde f(\gamma))\over d \gamma}\biggr),}
where $\chi^1(\gamma)$ can also be found in \NLOBFKLlf. As in \S 3,
this is a first order differential equation in $\gamma$, and it can be
solved in the same manner. In fact it is rather easier to
alter \NLOBFKL\ to the slightly different form 
\eqn\NLOBFKLalt{f(k^2,Q_0^2,\mu^2)= f^0(k^2,Q_0^2)
+\biggl({\bar\alpha_s(k^2) \over N}\biggr)
\int_0^{\infty}{dq^2\over q^2}(K^0(q^2,k^2)
-\alpha_s(\mu^2)K^1(q^2,k^2))f(q^2).}
This is identical to \NLOBFKL\ up to NNLO in $\alpha_s(\mu^2)$ and is
a common way for the NLO BFKL equation to be written since it makes
the solution easier. One must simply remember that the solution
obtained is only uniquely defined up to NLO in $\alpha_s(\mu^2)$ when
the coupling $\alpha_s(k^2)$ is expanded about $\alpha_s(\mu^2)$. 
If we take the
Mellin transformation of \NLOBFKLalt\ with respect to
$(k^2/\Lambda^2)$ we obtain 
\eqn\bfklrunnlo{{d \tilde f(\gamma,N)\over d \gamma}={d\tilde 
f^0(\gamma, Q_0^2) \over
d\gamma}-{1\over \bar\beta_0 N} (\chi_0(\gamma)-\alpha_s(\mu^2)\chi_1(\gamma))
\tilde f(\gamma, N),}
which is identical to \bfklrunii\ except for the NLO in $\alpha_s(\mu^2)$ 
correction to the
kernel. It can therefore be solved in exactly the same manner as this
previous equation (this would also be true for \nlomell), 
again obtaining a solution factorizing into
a $Q^2$-dependent part and a $Q_0^2$ dependent part. Each of these is
a contour integral and analogously to the previous treatment 
expanding about the saddle point
when performing the inverse Mellin transformation to $Q^2$- or
$Q_0^2$-space produces an
ordered series in $\alpha_s(\mu^2)$, as long as we also expand 
$\alpha_s(Q^2(Q_0^2))$ about $\alpha_s(\mu^2)$. 
This time the saddle point is at \ref\ciafanom{G. Camici and M. Ciafaloni
\PL \vyp{B386}{1996}{341}.}
\eqn\sadpointnlo{\gamma^{NLO,SP}(\bar\alpha_s/N)=\gamma^0(\bar\alpha_s/N)
-\beta_0\alpha_s\ln(Q^2(Q_0^2)/\mu^2){\partial \gamma^0(\bar\alpha_s/N)
\over \partial
(\ln(\alpha_s))} - \alpha_s{\chi_1(\gamma^0(\bar\alpha_s/N))
\over -\chi'_0(\gamma^0(\bar\alpha_s/N))}+\cdots,}
where
\eqn\wrongnloan{\alpha_s{\chi_1(\gamma^0(\bar\alpha_s/N))\over 
-\chi'_0(\gamma^0(\bar\alpha_s/N))} \equiv \alpha_s \gamma^1(\bar\alpha_s/N),}
is often called the NLO-BFKL anomalous dimension, and all other
corrections are beyond NLO in $\alpha_s$.\foot{For the remainder of this 
section
unless the argument is explicitly stated $\alpha_s\equiv \alpha_s(\mu^2)$.} 

Using the previous choice of input we can evaluate the two inverse 
transformations about the saddle point (we only need go 
further than the strict saddle point approximation when  
considering the $(1/\gamma)$ factor in the $Q^2$-dependent integrand
to obtain all results up to NLO accuracy - i.e. we use 
$1/(\gamma^0 +\gamma)^{-1}=1/\gamma^0 -\gamma/(\gamma^0)^2$). 
This gives a solution for
the gluon structure function of the 
form
\eqn\nlosol{\eqalign{{{\cal G}(Q^2,N)\over g_B(Q_0^2,N)}&
= {1 \over (\bar\alpha_s/N)
\gamma^0(-\chi'_0(\gamma^0))}\biggl(1-\alpha_s\biggl(-{\gamma^1\over \gamma^0}
+\gamma^1{\chi''_0(\gamma^0)\over -\chi'_0(\gamma^0)}+{\chi'_1(\gamma^0) \over
-\chi'_0(\gamma^0)}-\beta_0{\chi''_0(\gamma^0)\chi_0(\gamma^0)\over
2\gamma^0(-\chi'_0(\gamma^0))^2}\biggr)\cr
&\hskip -0.8in -\beta_0\alpha_s\ln(Q^2/\mu^2)
{\partial \gamma^0 \over \partial \ln(\alpha_s)}\biggl(
-{1\over \gamma^0}
+\half{\chi''_0(\gamma^0)\over -\chi'_0(\gamma^0)}\biggr) 
-\beta_0\alpha_s\ln(Q_0^2/\mu^2)\biggl(-1+
\half{\partial \gamma^0\over \partial \ln(\alpha_s)}
{\chi''_0(\gamma^0)\over -\chi'_0(\gamma^0)}\biggr)\biggr)\cr
&\times\exp\biggl(\int_{Q_0^2}^{Q^2} (\gamma^0
-\beta_0\alpha_s\ln(q^2/\mu^2){\partial \gamma^0
\over \partial
(\ln(\alpha_s))} - \alpha_s \gamma^1)\, d\ln q^2 \biggr).\cr}}
This allows us to determine the gluon coefficient function and gluon
anomalous dimension up to NLO in $\alpha_s(\mu^2)$, where the former
may be defined as the
value of \nlosol\ when $Q_0^2=Q^2$, and the latter is then determined by the
evolution of \nlosol\ with respect to $Q^2$ once the coefficient
function has been subtracted out,
i.e.
\eqn\nlocoeff{\eqalign{{\cal R}^{NLO}(\bar\alpha_s/N,Q^2/\mu^2)
=&{1 \over (\bar\alpha_s/N)
\gamma^0(-\chi'_0(\gamma^0))}\biggl(1-\alpha_s\biggl(-{\gamma^1\over \gamma^0}
+\gamma^1{\chi''_0(\gamma^0)\over -\chi'
_0(\gamma^0)}+{\chi'_1(\gamma^0) \over
-\chi'_0(\gamma^0)}\biggr)\cr
&\hskip -0.5in+\alpha_s\beta_0\biggr({\chi''_0(\gamma^0)\chi_0(\gamma^0)\over
2\gamma^0(-\chi'_0(\gamma^0))^2}-\ln(Q^2/\mu^2)\biggl(-1+
{\partial \gamma^0 \over \partial \ln(\alpha_s)}\biggl(
-{1\over \gamma^0}
+{\chi''_0(\gamma^0)\over -\chi'_0(\gamma^0)}\biggr)\biggr)\biggr)
\biggr).\cr}}
and 
\eqn\nloanominc{\gamma^{NLO}
(\bar\alpha_s/N, Q^2/\mu^2)=\gamma^0
-\beta_0\alpha_s\ln(Q^2/\mu^2){\partial \gamma^0
\over \partial
(\ln(\alpha_s))} - \alpha_s \gamma^1 + \biggl(-1+
\half{\partial \gamma^0 \over \partial \ln(\alpha_s)}
{\chi''_0(\gamma^0)\over -\chi'_0(\gamma^0)}\biggr).}
It is gratifying, though necessary, that in both cases the LO results from 
\soliv\ are are reproduced, and the terms $\sim \beta_0\alpha_s\ln(Q^2/\mu^2)$
are consistent with the renormalization group. (Note that 
$-\alpha_s\gamma^1(\bar\alpha_s/N)$ is not really the NLO correction to the
anomalous dimension in this scheme - it is actually quite similar to the 
$\overline{\hbox{\rm MS}}$ factorization scheme anomalous dimension.) 
Nevertheless, both of these quantities are
dependent on our choice of input and factorization scheme, and do not
contain any real physics.    

The only physically unambiguous quantity which may
be extracted is the effective anomalous dimension defined by \evol:
\eqn\nloanom{\Gamma(N,Q^2/\mu^2)=\gamma^0-
\beta_0\alpha_s\biggl({\partial \gamma^0\over
\partial \ln(\alpha_s)}\ln(Q^2/\mu^2)+  
{\partial \gamma^0 \over \partial\ln(\alpha_s)}
\biggl({-\chi''(\gamma^0) \over
2\chi'(\gamma^0)}-{1\over \gamma^0}\biggr)\biggr)-\alpha_s\gamma^1.}
The second term on the right corresponds to the NLO in $\alpha_s$
contributions previously accounted for 
when considering the running coupling,
while the third gives the additional NLO corrections. 
By examining the part of $\gamma^1$ which depends on $\beta_0$ we can check
whether at NLO at least the previous assumption about the manner in which to 
treat running coupling effects was correct, i.e. can see whether these 
do give the dominant contribution at NLO, or whether the conformal parts of 
$\gamma^1$ are more important. 

One can study the terms in \nloanom\ by finding the explicit form of each
as a power series in $\bar \alpha_s/N$. However, in the small $x$ limit we can 
examine the form of the singularities in the $N$-plane, i.e. the limit
of each of the terms as $\gamma^0 \to \half$ and $N\to \lambda$. Using the
well publicized fact that $\chi_1(\half)=4\ln2\times 6.3$ for $4$ flavors
and in the $\overline{\hbox{\rm MS}}$ renormalization scheme, 
and taking the inverse transformation back to $x$-space of \nloanom,
we obtain
\eqn\nlospli{xP(x,Q^2) =\bar\alpha_s\exp(\lambda\xi)\biggl(
{0.068 \over (\bar\alpha_s\xi)^{3/2}}-\beta_0\alpha_s\biggl(\biggl({0.188 \over
(\bar\alpha_s\xi)^{\half}}\biggr)\ln(Q^2/\mu^2)+0.69\biggr)-
\alpha_s\biggl({1.18 \over
(\bar\alpha_s\xi)^{\half}}\biggr)\biggr).}
Hence, the last term, although numerically large, is subleading to the effects
due to the running of the coupling we have previously considered, being a 
power of $(\bar\alpha_s\xi)^{\half}$ smaller. 
However, now we can be a little more
systematic. Examining the full NLO correction $\chi_1(\gamma)$, presented in
eq. (14) in \NLOBFKLlf, we see that there  are contributions which may be
interpreted as being due to the running of the coupling. These are 
$\half\beta_0(\chi^2(\gamma)+\chi'(\gamma))$  
and $-(5/3)\beta_0\chi(\gamma)$, 
coming from the NLO correction to the reggeon-reggeon-gluon vertex 
and the purely virtual terms respectively.\foot{It does not seem certain
whether or not the second of these terms should be included as a running 
coupling effect or not. As will become clear below this is only relevent for 
the scale choice at high $x$ where other considerations from large $x$ terms
come into play also.}
We imagine that these should be moved out of $\gamma^1$ in \nloanom\ and put
into the $\beta_0$-dependent part of the NLO correction. Doing this changes
\nlospli\ to
\eqn\nlosplii{xP(x,Q^2) =\bar\alpha_s \exp(\lambda\xi)\biggl(
{0.068 \over (\bar\alpha_s\xi)^{3/2}}-\beta_0\alpha_s\biggl(\biggl({0.188 
\ln(Q^2/\mu^2) -0.05\over
(\bar\alpha_s\xi)^{\half}}\biggr)+0.69\biggr)-
\alpha_s\biggl({1.23 \over
(\bar\alpha_s\xi)^{\half}}\biggr)\biggr).}
Therefore, not only is this additional NLO correction due to the
running of the 
coupling numerically very small, but it is also subleading at small
$x$ to the 
terms we have already considered.\foot{Not including the  $-(5/3)\beta_0
\chi(\gamma)$ term would simply lead to $-0.05$ becoming $0.26$ and $1.23$
subsequently becoming $0.92$.}
Choosing the renormalization scale 
$\mu$ by setting the $\beta_0$ dependent term to zero\foot{I choose  
$\alpha_s$ to be $\alpha_s(Q^2)$ rather than $\alpha_s(\mu^2)$ when doing this.
The two are of course equivalent up to higher order corrections, but the 
results of previous sections suggest that this is the appropriate choice.},
we obtain a very minor correction to our previous choice of scale
for the limit $x \to 0$, i.e.
\eqn\mincorscale{\ln(Q^2/\Lambda^2)\to \ln(Q^2/\Lambda^2) -0.26+
{(56\zeta(3)\pi)^{\half}\over 4}(\bar\alpha_s(Q^2)\xi)^{\half},}
where in fact there should really an additional constant on the right in the 
above equation
due to subleading corrections in \expscale\ that I have ignored.
The constant on the right of \mincorscale\ is also renormalization 
scheme dependent, though the dominant $3.63(\bar\alpha_s(Q^2)\xi)^{\half}$
term is not. 
  
We can also solve the equation for the scale exactly rather than in 
the small $x$ limit.
Putting our additional terms into the definition of the running coupling
dependent NLO splitting function, we compare with the previous 
$\hat p^1(\bar\alpha_s(Q^2)\xi)$ in \fig\nlocomp{Comparison of the exact NLO
$\beta_0$-dependent splitting function $\hat
p^1(\bar\alpha_s(Q^2)\xi)$ including the corrections from $\gamma^1$
with the value of $\hat
p^1(\bar\alpha_s(Q^2)\xi)$ obtained using the assumption in \S 4. Both terms
are weighted by $(\exp(2\alpha_s(Q^2)\xi))^{-1}$ for ease of
comparison.}. We see that indeed the
corrected $\hat p^1(\bar\alpha_s(Q^2)\xi)$ is slightly smaller
than the original for $\bar \alpha_s(Q^2)\xi \geq 1$ but is 
different at higher $x$, implying a different scale choice here
to that in \S 4. Of course, at these higher values of $x$ the 
differences are not too
important since, as already mentioned, the scale changes are
small here, and there will be interference with other effects from the
order by order in $\alpha_s$ expansion. 

Hence, we find that at NLO our previous assumption about the $-\beta_0
\alpha_s(\mu^2)\ln(k^2/\mu^2)$ term (which had to be present) being the
dominant contribution associated with the running coupling is very
well justified. This gives us confidence, if not a proof, that the
approach taken in the previous sections, i.e. that the     
$(-\beta_0\alpha_s(\mu^2)\ln(k^2/\mu^2))^n$ terms are the dominant
contribution from the running of the coupling at all orders is roughly
correct. Consequently, this full NLO result also supports the
hypothesis that the LO running coupling effects can be taken account of 
simply by using the $x$-dependent scale choice, determined by the BLM
prescription, in the LO expression for the effective splitting
function. 

Before considering the details of the NLO corrections to the kernel
which are not associated with the running of the coupling let us
reconsider the NLO BFKL equation. Given the above results it seems very
unlikely that the NLO BFKL equation as written in \NLOBFKLalt\ is
will be a good representation of the real physics since
the overall power of the coupling is allowed to run with $k^2$ while
that associated with the NLO kernel is fixed at
$\mu^2$. Bearing in mind that letting the coupling run in the LO
equation leads to such dramatic effects, and that at higher orders
there will definitely be the logs in $(k^2/\mu^2)$ associated with the
running of this additional factor of $\alpha_s(\mu^2)$ (with what now seem
likely to be small corrections) it seems most appropriate to write the
NLO BFKL equation with running coupling as 
\eqn\NLOBFKLalti{f(k^2,Q_0^2,\mu^2)= f^0(k^2,Q_0^2)
+\biggl({\bar\alpha_s(k^2) \over N}\biggr)
\int_0^{\infty}{dq^2\over q^2}(K^0(q^2,k^2)
-\alpha_s(k^2)K^1(q^2,k^2))f(q^2),}
if attempting to find a complete solution, as proposed in \ciafpcol. 
Strictly speaking 
$\alpha_s(k^2)$ should then be the two-loop running coupling, but this
will make the equation very complicated.  I will just use the one-loop
coupling which leads to a 2nd order differential equation in
$\gamma$-space
\eqn\bfklrunnlotot{{d^2 \tilde f(\gamma,N)\over d \gamma^2}={d^2\tilde 
f^0(\gamma, Q_0^2) \over
d\gamma^2}-{1\over \bar\beta_0 N} {d (\chi_0(\gamma)\tilde 
f(\gamma,N))\over d\gamma}-{\pi\over 3\bar\beta^2_0 N}\chi_1(\gamma)
\tilde f(\gamma, N).}

This can be solved in a very similar way to the approach in \S 4,
i.e. at leading twist it factorizes into the same form as \solruniv\
\eqn\solrunnloi{{\cal G}^{NLO}(N,Q^2)=a^{NLO}(Q^2,N)
b^{NLO}(Q_0^2,N)g_B(Q_0^2,N),}
where
\eqn\solrunnloii{a^{NLO}(Q^2,N)={1\over 2\pi i}
\int_{\half -i\infty}^{\half+i\infty}
{1\over \gamma}
\exp(\gamma\ln(Q^2/\Lambda^2)-X_{NLO}(\gamma,N)/(\bar\beta_0 N))
d\gamma.}
However, $X_{NLO}(\gamma,N)$ is rather more
complicated than the previous $X(\gamma)$
It can be expressed in the form 
\eqn\solrunnloiii{X_{NLO}(\gamma,N)=\int_{\half}^{\gamma}
\chi_{NLO}(\hat\gamma,N)d\hat\gamma,}
where $\chi_{NLO}(\gamma,N)$ can be written as a power series in $N$ 
beginning at zeroth order with $\chi_0(\gamma)$. As seen in \ciafpcol,
though here ignoring any resummations in $N$, the explicit form is 
\eqn\solrunnloiv{\chi_{NLO}(\gamma,N) =
\chi_0(\gamma)-N{\chi_1(\gamma) \over \chi_0(\gamma)} +{N^2 \over
\chi_0}\biggl(-\biggl({\chi_1(\gamma)\over
\chi_0(\gamma)}\biggr)^2-\beta_0 \biggl({\chi_1(\gamma)\over 
\chi_0(\gamma)}\biggr)'\biggr) +\cdots,}
where $\chi_2(\gamma)$ would also appear at order $N^2$ if I had included
it.  

It is now possible to obtain some general and rather specific results
using \bfklrunnlotot. Putting \solrunnloiv\ into \solrunnloii\ we note
that the leading singularities in $\gamma$ and $N$ are still both at
$0$, and thus there is still no true powerlike growth.
Furthermore, the singularity at $N=0$ is not affected by any of the
additional terms in \solrunnloiv\ beyond $\chi_0(\gamma)$ since in 
the exponent in \solrunnloii\ the ${\cal O}(N)$ term leads to a
constant as $N\to 0$ and all higher order terms vanish in this limit. Hence,
none of these terms should affect the solutions in the limit $x\to 0$,
except that the ${\cal O}(N)$ term should affect the overall
normalization, and we still expect small $x$ solutions 
$\sim \exp((\xi)^{\half})$ with the exponent the same as in the LO
case. Hence, higher order corrections to the BFKL equation should be 
very subleading when calculating physical quantities. This implies
that the scale for the coupling in higher order corrections should be
of the same type as at LO, i.e. falling with $x$. 

It is also possible to be more quantitative. \solrunnloii\ can be
solved using the same techniques as in \S 4 - expanding about the
saddle point leads to an ordered expansion in $\alpha_s(Q^2)$. 
Using \solrunnloiv\ it is easy to find that the
saddle point is now at 
\eqn\sadpointnlorun{\gamma^{SPNLO}(\bar\alpha_s(Q^2)/N)=
\gamma^0(\bar\alpha_s(Q^2)/N)
- \alpha_s(Q^2){\chi_1(\gamma^0(\bar\alpha_s(Q^2)/N))
\over \chi'_0(\gamma^0(\bar\alpha_s(Q^2)/N))}+{\cal O}(\alpha^2_s(Q^2)).}
Expanding as in \sadii\ one finds the saddle point solution
\eqn\sadiiinlo{\eqalign{a^{NLOSP}(Q^2,N)= &{1 \over \gamma^{SPNLO}
(\bar\alpha_s(Q^2)/N,N)(-(\chi'_{NLO}(
\gamma^{SPNLO}(\bar\alpha_s(Q^2)/N),N)))^{\half}}\cr
&\hskip 2in \exp\biggl(\int^{Q^2}\gamma^{SPNLO}(\bar
\alpha_s(q^2)/N) d\ln q^2\biggr).\cr}}
Further corrections can be calculated as in \S 4. However, this
expression contains some interesting information - the dominant
contribution to the running coupling corrections to the conformal part
of the NLO effective splitting function. Calculating $\Gamma(N,Q^2)$ as
a power series in $\alpha_s(Q^2)$ and transforming to $x$ space one
recovers all the contributions to the splitting functions in
\S 4. One also obtains the term $-\alpha_s(Q^2)p^{1,conf}
(\bar\alpha_s(Q^2)\xi)$ which is the transformation of 
$-\alpha_s(Q^2)\gamma^1(\alpha_s(Q^2)/N)$
(with the $\beta_0$-dependent terms extracted), and contributions to
the $\beta_0\alpha_s^2(Q^2)p(\bar\alpha_s(Q^2)\xi)$ splitting
function. This latter term provides the scale appropriate to use in the
NLO conformal splitting function using the BLM prescription at 
NLO \ref\blmnlo{S.J. Brodsky and 
H.J. Lu, \PR \vyp{D51}{1995}{3652}.}. This usually gives different
choices for the LO and NLO scales, which could be particularly
important in this case where the scale choice is so important.  
 
Calculating $\Gamma(N,Q^2)$ from \sadiiinlo\ the NLO conformal
contribution 
\eqn\nloconfan{-\alpha_s(Q^2)\gamma^1(\alpha_s(Q^2)/N)\equiv  
- \alpha_s(Q^2){\chi_1(\gamma^0(\bar\alpha_s(Q^2)/N))
\over -\chi'_0(\gamma^0(\bar\alpha_s(Q^2)/N))}}
comes from the argument of the exponential term. The leading
contribution
to the $\beta_0$-dependent correction to this comes from the expansion 
of 
\eqn\leadcont{
\biggl(-\chi'_0\biggl(\gamma^0(\bar\alpha_s(Q^2)/N)
- \alpha_s(Q^2)
{\chi_1(\gamma^0(\bar\alpha_s(Q^2)/N))
\over -\chi'_0(\gamma^0(\bar\alpha_s(Q^2)/N))}\biggr)\biggr)^{-\half},}
to order $\alpha_s(Q^2)$ which in $\Gamma(N,Q^2)$ leads to the term
\eqn\nloevolnlo{\beta_0\alpha^2_s(Q^2)
{d \gamma^0(\bar\alpha_s(Q^2)/N) \over d\ln(\alpha_s(Q^2))}
\biggl({(\chi''_0(\gamma^0(\bar\alpha_s(Q^2)/N)))^2 
\chi_1(\gamma^0(\bar\alpha_s(Q^2)/N)) \over
(-\chi'_0(\gamma^0(\bar\alpha_s(Q^2)/N)))^3}\biggr).}
It is easy to check that all other terms of ${\cal
O}(\beta_0\alpha^2_s(Q^2))$ are less
divergent as $N\to \lambda(Q^2)$ than
this one, including the contributions due to the $\beta_0$-dependent term
appearing explicitly in \solrunnloiv, which are very
subleading. Similarly, the contributions from the unknown $\chi_2(\gamma)$
will be very subleading unless $\chi_2(\gamma)$ is rather singular at
$\gamma =\half$. Taking the $\ln(Q^2)$-derivative of \nloconfan\ and
transforming this and \nloevolnlo\ to $x$ space one may find the 
scale for the NLO splitting function in the same way that the scale 
for the LO splitting function was found in \S 4. However,
comparing \nloconfan\ and \nloevolnlo\ with the terms in \nloevol\ one 
notices some similarities. These are not accidental, and a careful 
analysis  following the lines of \losplit\ to \nlosplit\ leads to
exactly the same result as at LO - the scale appropriate to the NLO
conformal splitting function is given by
\eqn\repeat{\ln(Q^2/\Lambda^2)\to \ln (\tilde Q^2/\Lambda^2) = 
\ln(Q^2/\Lambda^2) +
{(56\zeta(3)\pi)^{\half}\over 4}(\bar\alpha_s(Q^2)\xi)^{\half}.}
This exact equality was not at all guaranteed and is a remarkable 
result, implying the universality of this
scale choice at all orders. It is also renormalization scheme
independent, like the asymptotic form of the LO scale choice.
It is undoubtedly true that the LO scale and
the NLO scale will differ for finite $x$, this depending on the
unknown NNLO kernel, but it shows that the asymptotic results are very
simple and 
perturbation theory ought to be particularly convergent at small $x$. 
The NLO scale also matches well with the qualitative predictions
obtained from consideration of the singularity structure of the full
solution, as we will see below.

Using this scale at NLO we can investigate the precise effects of the NLO
corrections not associated with the running coupling, the so-called
conformal contributions.    
To begin with I simply remove the $\beta_0$-dependent terms from 
\nlosplii\ obtaining
\eqn\nlosplicon{xP(x,Q^2) =\alpha_s \exp(\lambda\xi)\biggl(
{0.068 \over (\bar\alpha_s\xi)^{3/2}}-
\alpha_s\biggl({1.23 \over(\bar\alpha_s\xi)^{\half}}\biggr)\biggr).}
Therefore, considering $\alpha_s$ as a constant for the moment, we see
that the NLO correction is both numerically large, and enhanced by a
power of $\bar\alpha^2_s\xi$ compared to the LO. This latter point is
really expected. Consider a leading order result of the form
$\exp(A\bar\alpha_s \xi)$. When we go to NLO the coupling constant
$\alpha_s$ becomes a renormalization scheme dependent quantity,
uncertain by ${\cal O}(\alpha_s^2)$. In order to be consistent with
the renormalization group and produce a result which is independent of
renormalization scheme up to higher orders the form of the full
solution must be
$\exp((A\bar\alpha_s+B\bar\alpha_s^2+\cdots)\xi)$,
where $B$ is scheme dependent. Expanding this about the
LO solution we get $\exp(A\bar\alpha_s\xi)(1+B\bar\alpha^2_s\xi+\cdots)$,
i.e. the NLO correction is indeed a power of $\bar\alpha_s^2\xi$ times the LO
result, exactly what we see in \nlosplicon. From this
argument it is clear that the NLO correction should be exponentiated,
and we obtain
\eqn\nlospliconexp{xP(x,Q^2) =\bar\alpha_s 
{0.068 \over (\bar\alpha_s\xi)^{3/2}}\exp(\lambda\xi(1-6.5\bar\alpha_s)),}
i.e. we obtain (slightly altered due to the removal of the
$\beta_0$-dependent term) the publicized correction to the
powerlike behaviour. 

However, we know that
$\alpha_s$ is not a constant, but runs according to our scale
choice at both LO and NLO. Indeed, the renormalization group argument 
above shows that
the NLO terms in \nlospli\ which behave like $-0.69\bar\alpha_s\beta_0\alpha_s
\exp(\lambda\xi)$ are not of the form we would naturally expect for the
NLO corrections, i.e. are not just a power of $\alpha_s$ higher, do
not represent the order of renormalization scheme uncertainty, and are 
not really subleading. Resumming by absorbing them into the definition 
of $\alpha_s$ seems the only
sensible thing to do. Doing this and using the scale choice 
\expscale\ in the
small $x$ limit in the expression \nlospliconexp\ in both the LO and NLO
parts gives 
\eqn\nloslpiconscal{xP(x,Q^2) \propto{1\over  (\alpha_s(Q^2)\xi)^{\half}
(\xi/\alpha_s(Q^2))^{3/4}}\exp(1.14(\xi/\alpha_s(Q^2))^{\half}-
3.0/\alpha_s(Q^2)).}
Therefore, it is only the LO part which gives the $x$ dependence in
this limit. The NLO part gives a $Q^2$-dependent normalization change,
which can admittedly be large (though using the $x \to 0$ limit of
\expscale\ tends to exaggerate the size of this at finite
$x$), as expected from the singularity structure of the solution of
the full NLO BFKL equation. 
Hence, using this scale choice the log of $xP(x,Q^2)$ is very insensitive to
NLO corrections at small $x$, and we would expect the NNLO corrections
to $\to 0$ as $x\to 0$. \foot{This result for the splitting
functions as $x \to 0$ is $xP(x,Q^2)\sim
\exp(A_{LO}(\xi/\alpha_s(Q^2))^{\half} -B_{NLO}/\alpha_s(Q^2))$
where $A_{LO}$ is renormalization scheme independent, 
$B_{NLO}$ is scheme dependent and higher order corrections are
claimed to be negligible. The apparent scheme dependence can be eliminated
by noting that the leading order result assumed $\ln(\tilde Q^2/\Lambda^2) = 
3.63(\bar\alpha_s(Q^2)\xi)^{\half}$  as $x\to 0$. Including the full
$\ln(\tilde Q^2/\Lambda^2) =\ln(Q^2/\Lambda^2) +B_{LO}+
3.63(\bar\alpha_s(Q^2)\xi)^{\half}$, where $B_{LO}$ is renormalization 
scheme dependent, leads to $xP(x,Q^2)\sim
\exp(A_{LO}(\xi/\alpha_s(Q^2))^{\half}+ C_{LO}/(\alpha_s^2(Q^2))+
(B_{LO}-B_{NLO})/\alpha_s(Q^2))$
where $C_{LO}$ and $B_{LO}-B_{NLO}$ are scheme -independent.}

Therefore, I conclude that the remaining NLO corrections, after running
coupling effects have been absorbed into the LO expression, are made far less
significant by the effective scale used, which has been shown to be
the same for LO and NLO. However, they are still potentially
important at small $x$. As far as comparison with experiment is
concerned the interesting question is whether these NLO corrections are
significant within the current range of data available. In order to
answer this question it is probably better to adopt a more
sophisticated procedure, and look at the evolution not of some
hypothetical gluon structure function, but of the true physical structure
functions. 

\newsec{Small $x$ Structure Functions.}

The previous sections have all considered the calculation of the gluon
structure function obtained by integrating the solution of the BFKL equation
up to the virtuality $Q^2$. Of course, this gluon structure function
is not a real physical quantity, though it does, as we shall
see, contain most of the essential information for physical quantities
for asymptotically small $x$. However, we would like to see precisely
how the results in the previous sections apply to real physical
quantities, and how universal they are. 

The generalization of the previous results to real physical scattering
processes is quite straightforward. Instead of integrating the upper
leg of the gluon Green's function from zero up to $Q^2$ we perform the
convolution of the Green's function with the scattering cross-section
for a probe of virtuality $Q^2$ with a gluon of virtuality $k^2$ \kti. i.e.
\gluondef\ is replaced by
\eqn\strcfun{{\cal F}_{i}(Q^2,N)=\int_{0}^{\infty}{dk^2\over k^2} 
\sigma_{i,g}(k^2/Q^2, \alpha_s(\mu^2)) f(N,k^2,Q_0^2)
g_B(N,Q_0^2).}
Currently the relevent $\sigma_{i,g}(k^2/Q^2, \alpha_s(\mu^2))$ are known at 
lowest nontrivial order for a number of quantities. This is order $\alpha_s$
for $F_L(x,Q^2)$ for both massive \ref\cat{S. Catani, Proceedings of DIS 96, 
Rome, April 1996, p. 165, {tt hep-ph/9608310}\semi S.Catani, 
\ZP \vyp{C75}{1997}{665}.} 
and massless quarks \ref\cathaut{S. Catani and F. Hautmann, \PL  
\vyp{B315}{1993}{157}; \NP  \vyp{B427}{1994}{475}.}, $F_2(x,Q^2)$ for massive 
quarks \kti\ and $(\partial F_2(x,Q^2)/\partial \ln Q^2)$ for massless quarks
\cathaut. For massless quarks the lowest order result for $F_2(x,Q^2)$ is 
zeroth order in $\alpha_s$ and is infrared divergent, 
representing the unknown nonperturbative quark distribution function.
None of the cross-sections are known beyond leading order, but all diagrams 
accounting for the running coupling corrections at NLO for the structure
functions are contained within the NLO BFKL equation. 

Taking the Mellin transformation of \strcfun\ with respect to 
$(Q^2/\Lambda^2)$ leads to the simple expression
\eqn\strcfunmell{\tilde {\cal F}_{i}(\gamma,N)= 
\alpha_s h_{i,g}(\gamma) \tilde {\cal G}(\gamma,N),}
where as before $\tilde {\cal G}(\gamma,N)=\tilde f(\gamma,N,Q_0^2)
g_B(N,Q_0^2)/\gamma$, and 
$h_{i,g}(\gamma)$ is a function of $\gamma$ which is finite 
at $\gamma=0$ and $\gamma =1/2$. 
Using the appropriate expression for $\tilde {\cal G}(\gamma,N)$
the inverse Mellin transformation may be performed in the 
same manner as before in order to give the moment space structure 
functions - considering the running coupling constant BFKL equation,
either LO or NLO, expanding about the same saddle points leads to an
ordered solution in $\alpha_s$. Let us examine the simple case of
${\cal F}_L(N,Q^2)$ with massless quarks only. 
As with the gluon structure function it is impossible to 
actually predict this function due to the unknown $g_B(N,Q_0^2)$
and due to the need to regularize the BFKL equation when using the running 
coupling. 
However, the previous leading twist factorization into an incalculable 
$Q_0^2$-dependent function and a calculable $Q^2$-dependent function 
also applies in the same way. 
The function $h_{L,g}(\gamma)$ is entirely associated with the 
latter and does not alter the previous properties for the case of the 
gluon - the $Q^2$-dependent function is a finite unambiguous quantity 
with a Mellin transformation having leading singularities at 
$\gamma=0$ and $N=0$. 

Hence, as in the case of the gluon structure function the entirely 
perturbative calculable quantity to consider is 
\eqn\physlong{\Gamma_{LL}(Q^2,N)={\partial \ln({\cal F}_L(N,Q^2))\over \partial
\ln Q^2}.}  
This can be calculated for the case of the running coupling and the LO 
BFKL equation as in \S 4, with all general results being the same as in 
this previous case, i.e. one obtains an oscillating series in 
$\beta_0\alpha_s(Q^2)$ and a very similar apparent scale choice, as we 
will see below. The 
changes brought about by using the NLO BFKL equation with running coupling
are also much the same as when considering the gluon. As stated, to get a 
full solution one should use the NLO BFKL equation in the 
form \NLOBFKLalti. Being instead entirely systematic one may use \NLOBFKLalt, 
and examine the results only up to NLO in $\alpha_s(\mu^2)$. Doing
this one calculates the analogues of \nloanominc\ and
\nlocoeff. The latter is unchanged while the former is altered by the
presence of $h_{L,g}(\gamma)$ into a different coefficient function
${\cal C}_L^{NLO}(\bar\alpha_s/N,Q^2/\mu^2)$. The evaluation of this
complete NLO coefficient function is not yet possible due to the
absence of the NLO correction to $\sigma_{L,g}(k^2/Q^2,
\alpha_s(\mu^2))$. However, in order to calculate the NLO physical anomalous 
dimension $\Gamma^{NLO}_{LL}(N,Q^2/\mu^2)$, the analogue of \nloanom, 
one needs only
the NLO part of ${\cal C}_L^{NLO}(\bar\alpha_s/N,Q^2/\mu^2)$ containing
$\ln(Q^2/\mu^2)$, which is really provided by the LO expression via the 
renormalization group.\foot{Equivalently one can use the formulae for
the physical anomalous dimensions  describing the evolution of structure 
functions in terms of themselves, rather than unphysical partons and 
coefficient functions, given in \cat.} Explicitly one obtains  
\eqn\nloanomll{\Gamma_{LL}(N,Q^2/\mu^2)=\gamma^0-
\beta_0\alpha_s\biggl({\partial \gamma^0\over
\partial \ln(\alpha_s)}\ln(Q^2/\mu^2)+  
{\partial \gamma^0 \over \partial\ln(\alpha_s)}
\biggl({-\chi''(\gamma^0) \over
2\chi'(\gamma^0)}-{1\over \gamma^0}+h'_{L,g}(\gamma^0)\biggr)\biggr)-
\alpha_s\gamma^1.}
Hence, the conformal part of $\Gamma_{LL}(N,Q^2/\mu^2)$ is identical
to that of $\Gamma(N,Q^2/\mu^2)$, but there is a modification to the term
determining the scale. In fact, the additional term, $h'_{L,g}(\gamma^0)$,
is a constant at $\gamma^0=\half$, and as such it only contributes
insignificantly as $x\to 0$: the asymptotic scale is dominated by
${\partial \gamma^0 \over \partial\ln(\alpha_s)}
\biggl({-\chi''(\gamma^0) \over
2\chi'(\gamma^0)}\biggr)$ and is identical to the 
choice already presented for the gluon structure function. 
$h'_{L,g}(\gamma^0)$ is important at moderate $x$, however. 

Taking the
transformation of \nloanomll\ back to $x$-space and eliminating the
$\beta_0$-dependent part (including the terms in $\gamma^1$) 
by setting the scale leads to a precise
definition of the effective coupling constant to be used for the
evolution of $F_L(x,Q^2)$ within this expansion scheme. This is
presented as a function of $x$ for two choices of $Q^2$ in
\fig\flscale{The effective coupling constant for the physical
splitting function $P_{LL}(x,Q^2)$ for $N_F=4$ 
as a function of $x$ compared to the constant values at the
relevant values of $\ln(Q^2/\Lambda^2)$.}, 
and can be compared with the effective coupling for the gluon
structure function (without the $\beta_0$-dependent terms in $\gamma^1$)
in \gluscale. Clearly  in both cases the effect of the change in scale
is to reduce the small $x$ coupling, and the effect becomes more important as 
$Q^2$ decreases and the size of  
$\hat p^1(\bar \alpha_s(Q^2)\xi)$ becomes larger relative to 
$p^0(\bar \alpha_s(Q^2)\xi)$.
However, for $F_L(x,Q^2)$ the effective coupling at    
$x=0$ is larger than $\alpha_s(Q^2)$. This is mainly due to
the $-(5/3)\beta_0\chi(\gamma)$ term in $\gamma^1$, but is also influenced by
the first nontrivial term in the series expansion of
$h_{L,g}(\gamma^0)$ in powers of $(\bar \alpha_s/N)$ which is
negative. As $x$ decreases the effective coupling quickly decreases
also, and soon falls below that in \gluscale. This latter point is due to the 
$\half\beta_0(\chi^2(\gamma)+\chi'(\gamma))$ term in $\gamma^1$ and
the remainder of $h_{L,g}(\gamma)$ which both act to increase $\hat
p^1_{LL}(\bar \alpha_s\xi)$, and hence increase the scale for the coupling. 
At $x=10^{-5}$ the effective coupling for $F_L(x,Q^2)$ is noticeably
lower than that for the gluon, but as $x$ decreases even further the
effect of the additional terms becomes less and less important, and
the couplings converge.  

One can now be rather quantitative about the phenomenological effects
of the NLO BFKL equation and the choice of scale. Let us first make
the simple scale choice $\mu^2=Q^2$. In this case we may write the
physical splitting function  as 
\eqn\splitll{\eqalign{(x/\bar\alpha_s(Q^2))
P_{LL}(\bar\alpha_s(Q^2)\xi)&=
p^0_{LL}(\bar\alpha_s(Q^2)\xi)-
\beta_0\alpha_s(Q^2)
p^{1,\beta}_{LL}(\bar\alpha_s(Q^2)\xi)-\alpha_s(Q^2) 
p^{1,conf}_{LL}(\bar\alpha_s(Q^2)\xi)\cr
&\equiv 
p^0_{LL}(\bar\alpha_s(Q^2)\xi)-
\alpha_s(Q^2)p^{1,tot}_{LL}(\bar\alpha_s(Q^2)\xi),\cr}}
where each of the $p^i_{LL}(\bar\alpha_s(Q^2)\xi)$ may be written as a power
series of the form\foot{Actually $p^{1,conf}_{LL}$ has an additional
term $\propto \delta(1-x)/(\alpha_s(Q^2))$ which appears in the normal
one-loop physical structure function.}
\eqn\powser{p^i_{LL}(\bar\alpha_s(Q^2)\xi)=\sum_{0}^{\infty}a_n
{(\bar\alpha_s(Q^2)\xi)^{n}\over n!}.}
The coefficients for the power series of the various terms in
\splitll\ are shown in table 1. As one can see the coefficients for
all the $p^1_{LL}(\bar\alpha_s(Q^2)\xi)$ are generally much larger
than those for 
$p^0_{LL}(\bar\alpha_s(Q^2)\xi)$. 

Using the conventional choice of scale then
at leading order one would obtain the value of $(\partial
F_L(x,Q^2)/\partial\ln Q^2)$ by convoluting the first term on the
right of \splitll\ with $F_L(x,Q^2)$ itself. As an appropriate choice
of $F_L(x,Q^2)$ at a value of $(Q^2/\Lambda^2)=8$ ($Q^2\sim 40\Gev^2$)
I choose $F_L(x,Q^2)=(x/0.1)^{-0.3}\Theta(0.1-x)$. This is a
function with the approximate shape of $F_2(x,Q^2)$ at this $Q^2$ and
the $\Theta$-function is chosen as a crude model for the approximate 
$(1-x)^6$ fall-off at large $x$. The result for the evolution of
$F_L(x,Q^2)$ is shown in the upper of \fig\evol{The values of
$(\partial F_L(x,Q^2)/\ln(Q^2))$ using the resummed physical splitting
functions for an input of
$F_L(x,Q^2)=(x/0.1)^{-0.3}\Theta(0.1-x)$ at $\ln(Q^2/\Lambda^2)=8$ as a
function of $x$. The upper figure shows the LO and LO $+$ NLO results
for the conventional scale choice $Q^2=\mu^2$. The lower figure shows
the LO, LO $+$ NLO and LO $+$ exponentiated NLO results for the $x$
dependent scale choice in this paper.}. It increases very
quickly at small $x$ due to both the shape of $F_L(x,Q^2)$ and the
large splitting function at small $x$. Using the conventional scale
choice one would then find the NLO evolution by using the whole of
\splitll. The effect of adding in this very large negative contribution
to the physical splitting function is also shown in the upper of
fig. 7. As one can see the effects are dramatic, largely killing the
evolution for $x>0.0001$ and turning it sharply negative below
this.\foot{Similar behaviour was found for the gluon in a particular 
factorization scheme ($\overline{\hbox{\rm MS}}$) using an incomplete
calculation of the NLO anomalous dimension \ref\incomplete{J. Bl\"umlein and
A. Vogt, \PR \vyp{D57}{1998}{1}; \PR \vyp{D58}{1998}{014020}.}. Using the
complete anomalous dimension does not alter the qualitative results.} 
Indeed, the NLO correction is  nearly as large as the LO result for
$x\sim 0.001$, and becomes dominant as $x$ decreases below this: the 
perturbative solution is not at all stable.  
Also, although we do not have measurements of $F_L(x,Q^2)$ in this
range of $x$ and $Q^2$, similar behaviour would feed through to
$F_2(x,Q^2)$, and the NLO prediction is dramatically at odds with the 
experimental data. This is therefore a real physical example of the 
problems induced by the NLO BFKL equation, and is completely
independent of factorization schemes and hence totally unambiguous
(which is not the case for discussions of behaviour of the gluon
distribution in a given factorization scheme). 
As we go to lower $Q^2$ the coupling becomes stronger and the expected
shape of the structure function becomes flatter. Both lead to the NLO
corrections becoming even more important relative to the LO, and at
$Q^2\sim 10\Gev^2$ the NLO correction is larger than the LO for
essentially all $x$. So
we see that the conventional choice for the scale leads to disastrous
results.

Let us consider instead the BLM scale choice for $P_{LL}(x,Q^2)$. 
Absorbing $p^{1,\beta}_{LL}(\bar\alpha_s(Q^2)\xi)$ into the definition of
the scale changes \splitll\ to 
\eqn\splitllblm{(x/\bar\alpha_s(\tilde Q^2))P_{LL}(\bar\alpha_s(\tilde
Q^2)\xi)=
p^0_{LL}(\bar\alpha_s(\tilde Q^2)\xi)-\alpha_s(\tilde Q^2) 
p^{1,conf}_{LL}(\bar\alpha_s(\tilde Q^2)\xi),}
where, as I have already noted, the LO scale is only guaranteed to be 
exactly the same as that to use at NLO as $x\to 0$. (Using
\NLOBFKLalti\ it is easy to show that this is true for $F_L(x,Q^2)$ in
the same way as for the gluon - $h_{L,g}(\gamma)$ only introduces
subleading effects as in \nloanomll.)
The result of the evolution using the LO
splitting function is shown in the lower of \evol. It is a little
smaller at the lowest values of $x$ than for $Q^2=\mu^2$, but only by
$\sim 15\%$. This is because until we get to extremely small $x$ the
LO evolution is largely driven by the first term in the power series
of $p^0_{LL}(\bar\alpha_s(\tilde Q^2)\xi)$ due to the vanishing of the
second, third and fifth terms, and relatively small fourth and sixth terms. 
Hence, the decrease of the coupling is only felt as a single power (and
indeed there is an increase of the coupling for the highest values of $x$). The
discrepancy between the LO results will increase at lower values of
$x$. It will also increase as $Q^2$ gets smaller and/or as the structure
function becomes less steep. It is when we include the NLO corrections
that the more dramatic result is seen. The size of these now decreases
for two reasons: much of the NLO correction has vanished, having been
absorbed
into the definition of the scale\foot{For lowish order in the power
series the coefficients for $p^{1,\beta}_{LL}(\bar\alpha_s(\tilde Q^2)\xi)$
and $p^{1,conf}_{LL}(\bar\alpha_s(\tilde Q^2)\xi)$ are similar, but the
former begin to dominate at higher orders, i.e. lower $x$, and become
totally dominant as $n\to \infty$ ($x\to 0$) as demonstrated by the
asymptotic results in the last section.}, and the effective coupling
is now much smaller at small $x$. The result of including the NLO
corrections is seen in the lower of \evol. It is now a significant,
but by no means overwhelming effect. As argued in the previous
section renormalization scheme consistency implies that these NLO
effects should really be exponentiated. The result of such an
exponentiation is also shown in the lower of \evol. It is clearly not
dramatic, but does help the convergence of the perturbative
calculation. The exponentiation will become more important as $x \to
0$. Now that I use the BLM scale choice the coupling at small $x$ is far less
sensitive to $Q^2$ than for $\mu^2=Q^2$ and  the relative importance of the 
NLO corrections increases far less quickly as $Q^2$ decreases. 
As shown for the case of the gluon, at asymptotically small
$x$ the effective splitting function will behave like 
$\exp(1.14(\xi/\alpha_s(Q^2))^{\half})$ and the exponentiated NLO
corrections will lead to an $x$-independent multiplicative factor. This
factor is potentially quite large, however, and the NLO effects must
ultimately be treated to obtain the correct quantitative
results. Nevertheless, it appears as though the LO calculation with
the correct scale setting may be quite accurate in the current range
of $x$ and $Q^2$ probed by experiment.  

These results regarding $F_L(x,Q^2)$ seem very pleasing. However,
phenomenologically $F_2(x,Q^2)$ is far more important since this is
the quantity for which we have a great deal of data \ref\hone{H1 
collaboration: S. Aid {\it et al.},\NP  \vyp{B470}{1996}{3}\semi
H1 collaboration: C. Adloff {\it et al.}, \NP  \vyp{B497}{1997}{3}.}
\ref\zeus{ZEUS collaboration: M. Derrick {\it et al},
\ZP  \vyp{C69}{1996}{607}\semi ZEUS collaboration: M. Derrick {\it et al},
\ZP  \vyp{C72}{1996}{399}.}. One can calculate
$(\partial F_2(x,Q^2)/\partial \ln Q^2)$ in exactly the same way as 
$F_L(x,Q^2)$ simply by using the relevent 
$\sigma_{2,g}(k^2/Q^2,\alpha_s(\mu^2))$ which leads to
$h_{2,g}(\gamma)$ in Mellin space. Hence, in this case one obtains a
direct expression for the evolution of the structure function with
respect to $Q^2$, rather than for the structure function
itself. However, inverting the Mellin transformation 
it is easy to see that the expression for 
$(\partial {\cal F}_2(N,Q^2)/\partial \ln Q^2)$ is identical to that
for ${\cal F}_L(N,Q^2)$ 
up to the $h_{i,g}(\gamma)$-dependent effective coefficient function 
(or in this case anomalous 
dimension). Whereas ${\cal F}_L(N,Q^2)$ has a factor of 
${\cal C}_L^{NLO}(\bar\alpha_s/N,Q^2/\mu^2)$ up to NLO, 
$(\partial {\cal F}_2(N,Q^2)/\partial \ln Q^2)$ has a factor of 
$\gamma_2^{NLO}(\bar\alpha_s/N,Q^2/\mu^2)$. Hence, we can write the
physical evolution equation 
\eqn\physanomtwo{{\partial {\cal F}_2(N,Q^2)\over \partial \ln Q^2} = 
\Gamma_{2L}(Q^2,N) {\cal F}_L(N,Q^2),}
where $\Gamma_{2L}(N,Q^2)=\gamma_2^{NLO}(\bar\alpha_s/N,Q^2/\mu^2)
/{\cal C}_L^{NLO}
(\bar\alpha_s/N,Q^2/\mu^2)$.\foot{Again one can use the rules for
finding physical anomalous dimensions in \cat.}
In this expression all the unknown nonperturbative physics associated
with $(\partial {\cal F}_2(N,Q^2)/\partial \ln Q^2)$ and 
${\cal F}_L(N,Q^2)$ cancels
out to leave us an entirely perturbatively calculable physical
anomalous dimension depending only on $Q^2$, $N$ and at finite order
our choice of $\mu$.  
  
As with ${\cal C}_L^{NLO}(\bar\alpha_s/N,Q^2/\mu^2)$ we do not know the NLO
off-shell cross-section and hence cannot fully calculate
$\gamma_2^{NLO}(\bar\alpha_s/N,Q^2/\mu^2)$. Hence, we cannot calculate 
$\Gamma^{NLO}_{2L}(N,Q^2)$ fully. However, we do know all the the
effects at NLO due to the running of the coupling for both
${\cal C}_L^{NLO}(\bar\alpha_s/N,Q^2/\mu^2)$ and
$\gamma_2^{NLO}(\bar\alpha_s/N,Q^2/\mu^2)$ and can calculate
the NLO contribution to $\Gamma^{NLO}_{2L}(N,Q^2)$ due to the running of the
coupling and hence find the appropriate scale to use in the LO
expression. This is a straightforward, though rather lengthy
calculation using the NLO BFKL equation in the form \NLOBFKLalt, and
expanding the Mellin-space solutions about the saddle point for both  
$(\partial {\cal F}_2(N,Q^2)/\partial \ln Q^2)$ and ${\cal F}_L(N,Q^2)$
in order to
find the relevent parts of ${\cal C}_L^{NLO}(\bar\alpha_s/N,Q^2/\mu^2)$ and
$\gamma_2^{NLO}(\bar\alpha_s/N,Q^2/\mu^2)$. It results in the
relatively simple expression\foot{For reasons of simplicity I have 
previously defined 
$\Gamma_{2L}(N,Q^2/\mu^2)$ with an additional factor of $\alpha_s$ 
\ref\lorsc{R.S. Thorne, \PL \vyp{B392}{1997}{463}; \NP 
\vyp{B512}{1998}{323}.}. This
leads to no differences when calculating physical quantities.} 
\eqn\nloanomtwol{\eqalign{\Gamma_{2L}(N,Q^2/\mu^2)&={h_{2,g}(\gamma^0)
\over h_{L,g}(\gamma^0)}-
\beta_0\alpha_s{\partial \gamma^0\over
\partial \ln(\alpha_s)}\biggl({\partial (h_{2,g}(\gamma)
/h_{L,g}(\gamma))\over
\partial \gamma}\biggr)_{\gamma^0}
\ln(Q^2/\mu^2)\cr
&\hskip -0.8in -\beta_0\alpha_s{\partial \gamma^0 \over \partial\ln(\alpha_s)}
\biggl(\biggl({-\chi''(\gamma^0) \over
2\chi'(\gamma^0)}-{1\over \gamma^0}\biggr)\biggl({\partial (h_{2,g}(\gamma)
/h_{L,g}(\gamma))\over
\partial \gamma}\biggr)_{\gamma^0}+\half\biggl({h''_{2,g}(\gamma^0)\over 
h_{L,g}(\gamma_0)}-{h''_{L,g}(\gamma^0)h_{2,g}(\gamma^0)\over 
h^2_{L,g}(\gamma_0)}\biggr)\cr
&+\biggl({\partial (h_{2,g}(\gamma)
/h_{L,g}(\gamma))\over
\partial
\gamma}\biggr)_{\gamma^0}\biggl(\half\biggl(\chi(\gamma^0)+{\chi'(\gamma^0)
\over \chi(\gamma^0)}\biggr)-{5\over 3}\biggr)\biggr),\cr}}
where $h_{2,g}(\gamma)$ and $h_{L,g}(\gamma)$ can be found in
\cathaut, and $(h_{2,g}(\gamma)/h_{L,g}(\gamma))=\tilde
\Gamma^0_{2L}(\gamma) = (3/2)\gamma +(1-\gamma)^{-1}$.
As usual we can take the transformation back to
$x$-space. Using the naive scale $Q^2=\mu^2$ we obtain
\eqn\nlosplittwol{(x/\bar\alpha_s(Q^2))P_{2L}(\bar\alpha_s(Q^2),x)=
{\delta(1-x)/\bar\alpha_s(Q^2)}+p^0_{2L}(\bar\alpha_s(Q^2)\xi)-
\beta_0\alpha_s(Q^2)
p_{2L}^{1,\beta}(\bar\alpha_s(Q^2)\xi),}
where the $p^i_{2L}(\bar\alpha_s(Q^2)\xi)$ are power series of the form
\powser, and the LO physical splitting function has a zeroth order
term proportional to a $\delta$ function. As is now standard we can 
find the correct scale by eliminating all $\beta_0$-dependent NLO
terms. This is a
little more involved than the previous cases, but in the asymptotic
limit reduces to exactly the same result. As
$\gamma^0 \to \half$ if we keep only the most divergent part in the
third term on the right in \nloanomtwol\ then we have the condition
that the $x$-space version of 
\eqn\condtwol{{\partial \gamma^0\over
\partial \ln(\alpha_s)}\biggl({\partial (h_{2,g}(\gamma)
/h_{L,g}(\gamma))\over
\partial \gamma}\biggr)_{\gamma^0}
\ln(Q^2/\tilde Q^2)
+{\partial \gamma^0 \over \partial\ln(\alpha_s)}
\biggl({-\chi''(\gamma^0) \over
2\chi'(\gamma^0)}\biggr)\biggr({\partial (h_{2,g}(\gamma)
/h_{L,g}(\gamma))\over
\partial \gamma}\biggr)_{\gamma^0},}
must vanish. Since in this limit $\biggl({\partial (h_{2,g}(\gamma)
/h_{L,g}(\gamma))\over
\partial \gamma}\biggr)_{\gamma^0} \to 5/2$ this is precisely the same
condition as we found for the gluon and for ${\cal F}_L(N,Q^2)$, and we
obtain exactly the same asymptotic scale \expscale. Indeed, if we attach any
physical process to the top of the gluon ladder we will always obtain
solutions for physical quantities in the same manner: the physical
anomalous dimension or coefficient function will be determined from
the part of the solution which has
factorized, is $Q^2$-dependent, and is influenced by the diffusion into the
ultraviolet. Hence, we would always expect physical quantities to be
controlled by the same asymptotic scale. 

Being more precise we may
find the $x$-space version of \nloanomtwol\ as a power series in 
$\bar\alpha_s(Q^2) \xi$. The
coefficient functions for the known $p^i_{2L}(\bar\alpha_s\xi)$ are shown 
in table 2. Using these series we can solve exactly for the scale down to
some finite value of $x$. The effective coupling to be used when
calculating the small $x$ evolution of $F_2(x,Q^2)$ in terms of
$F_L(x,Q^2)$ is actually very similar to that for the evolution of
$F_L(x,Q^2)$ over the whole range of $x$. They become identical as
$x\to 0$, but are only slightly different even as $x\to 1$. 

If we examine the value of $(\partial F_2(x,Q^2)/\partial \ln Q^2)$ 
for the given input for $F_L(x,Q^2)$ using the LO physical anomalous
dimension we find that the decrease in going from the choice
$Q^2=\mu^2$ to the effective scale is a little larger than when
examining $(\partial F_L(x,Q^2)/\partial \ln Q^2)$. This is simply
because the terms in the power series for $p^0_{2L}(\bar\alpha_s\xi)$
are not as small as those for $p^0_{LL}(\bar\alpha_s\xi)$, and so
higher terms in the series, where powers of the coupling are used, are
proportionally more important. Since we do not actually know the value
of $p^{1,conf}_{2L}(\bar\alpha_s\xi)$ it is impossible to evaluate the NLO
effects, with or without the scale setting, but I imagine they are of
similar importance to the those for $F_L(x,Q^2)$. They will certainly
lead to the same 
general result, i.e. the LO expression $\sim 
\exp(1.14(\xi)/\alpha_s(Q^2))^{\half}$ as $x \to 0$ with the exponentiated 
NLO corrections leading to an $x$-independent multiplicative factor. 

I note that within this picture there is no way of predicting inputs
for structure functions (or partons) at some fixed $Q_I^2$. However,
since the evolution generates no true powerlike behaviour there may well be
no growth at $x\to 0$ stronger than the soft pomeron. I see no reason to 
believe the values for the intercepts calculated by putting some infrared 
cut-off on the BFKL equation for running coupling, which are both cut-off
method and scale dependent. 
However, at the sort of values of $x$ we consider in
practice, $x=10^{-2}\to 10^{-5}$, the perturbative evolution can generate 
a rise at small $x$
which appears to be like an effective power over this restricted range in
$x$. In broad terms this will not be dissimilar to that generated by
the NLO in $\alpha_s(Q^2)$ evolution, but will be different in
detail. Perhaps the best method for attempting to predict the shape of 
a structure function at a given input scale is to demand that the
general form of the structure functions are as insensitive to changes
in starting scale as possible \lorsc. In this way the inputs are
determined largely by the form of the evolution, and hence the 
effective physical splitting functions.
Since the small $x$ evolutions of $F_2(x,Q^2)$ and $F_L(x,Q^2)$ are 
related in a calculable manner this imposes a precise consistency requirement 
on the small $x$ inputs of the two. A more detailed study of study of
this would be interesting, though an obvious conclusion is that the
shape of $F_2(x,Q^2)$ and $F_L(x,Q^2)$ with $x$ should be roughly the
same at all $Q^2$ and hence at $Q_I^2$ (see below).    
 
\newsec{Phenomenological Consequences.}   

Armed with the small $x$ scale choices for the physical structure functions,
it is now possible to do a phenomenological analysis. The inclusion of
the input singlet quark distribution, or equivalently the inclusion of
$\Gamma_{L2}(N,Q^2)$ and $\Gamma_{22}(N,Q^2)$ is easy since at LO
these are related in a simple manner to $\Gamma_{LL}(N,Q^2)$ and 
$\Gamma_{2L}(N,Q^2)$ respectively \cat. Furthermore, they have only a small
effect. Much more important is the treatment of the LO terms in the
physical splitting functions which are less singular than $1/x$ as
$x\to 0$. As shown in \lorsc\ a full LO analysis should include all
such terms at lowest order in $\alpha_s$ as well as all terms in the
LO small $x$ expansions considered so far in this paper. A correct extension 
of \lorsc, which used the simple scale choice $Q^2=\mu^2$, would involve
the full LO, in $\alpha_s$ as well as $\ln(1/x)$, physical
splitting functions with the scale choice determined not only by the
NLO running coupling effects considered in this paper, but also by the
$x$-finite NLO in $\alpha_s$ running coupling effects. 

Consideration of the NLO in $\alpha_s$ running coupling effects leads to 
additional
important scale changes away from $Q^2=\mu^2$ at high $x$. The
evolution of the nonsinglet structure function $F^{NS}_2(x,Q^2)$ was
considered in \ref\wong{W.K. Wong \PR \vyp{D54}{1996}{1694}.} where 
it was found that the appropriate scale to use is 
\eqn\wongscale{\tilde Q^2 = Q^2{(1-x)\over x^2}k(x),}
where $k(x)$ is a relatively smooth function of $x$ from $0\to 1$,
$k(x)\approx 0.15$. 
Careful consideration shows that such a scale change (with some 
regularization as $x\to 1$) must be implemented at high $x$ for quark
driven processes, leading to a larger coupling and quicker evolution.
There are also nontrivial high $x$ effects in the
gluon driven processes due to the NLO in $\alpha_s$ running coupling
terms. This changes the detailed form of the effective coupling already
presented in \flscale\ for values of $x$ above approximately
$x=0.05$. For values
of $x$ below this the finite $x$ effects on the scale fall away
quickly. 

One particular consequence of including the full ${\cal O}(\alpha_s)$
effective splitting functions is that like $P_{gg}(z,Q^2)$, 
$P_{LL}(z,Q^2)$ actually
leads to a fall with $Q^2$ for high values of $z$, the rise
only setting in when the small $x$ terms become dominant. Hence, the
fact that the effective coupling for $P_{LL}(z,Q^2)$ is actually large at
high $z$ increases this negative contribution, whereas the smaller
coupling at small $z$ decreases the positive contribution, as we
already know. This means that, looking at the complete convolution leading 
to the evolution of $F_L(x,Q^2)$, the increased negative contribution at 
high $z$ leads to  the 
full scale-fixed LO evolution being reduced compared to the full $Q^2=\mu^2$ LO
evolution more than the consideration of small $x$ effects only in
\evol\ suggests. Inclusion of the high $x$ terms at NLO has precisely
the opposite effect: this time the positive contribution 
to the evolution from high $z$ due to
the ${\cal O}(\alpha_s^2)$ terms is enhanced, as well as the known
effect of the negative contribution from small $z$ being much reduced in
size. Hence, the negative NLO correction at small $x$ is 
significantly reduced
compared to that seen in \evol. Details will be shown in a future
paper \ref\thornei{R.S. Thorne, in preparation.}, but the apparent 
convergence of the perturbative expansion is
considerably better even than that implied in the previous section.      

I leave a full discussion of the implementation of a
full LO in $\alpha_s$ and $\ln(1/x)$ (denoted by LORSC) global fit  
using scale setting
in physical anomalous dimensions to a future paper.\foot{It is also 
necessary to
treat the heavy partons in a consistent manner. The way to do this in
the context of the full LO physical anomalous dimensions with
$Q^2=\mu^2$ was presented
briefly in \ref\thornedis{R.S. Thorne, Proceedings of DIS 98, 
Brussels, April 1998, p. 207, {\tt hep-ph/9805299}.}, 
and will be presented in more detail in a 
future paper \ref\thorneii{R.S. Thorne, in preparation.}.} Details of 
such a (slightly approximate) fit have already been briefly reported in 
\durhamwg, and here I report the most important consequences.

\medskip 

\noindent 1. Compared to the most recent NLO in $\alpha_s(Q^2)$ global
fit \ref\mrst{A.D. Martin, R.G. Roberts, W.J. Stirling and R.S. Thorne,
{\it Eur. Jour. Phys.} \vyp{C4}{1998}{463}.} the quality of the 
$\chi^2$ is improved from 1511 to 1339 for 1330
structure function data points. (Constraints from non-structure
function data, e.g. prompt-photon, Drell-Yan {\it etc.} at high $x$ are
imposed in the same manner for both.) A breakdown of the $\chi^2$ 
for each experiment is shown in table. 3. This extremely statistically 
significant 
improvement is achieved in all regions of $x$ and
$Q^2$ - the scale choice \wongscale\ helping at high $x$ and the
resummation of $\bar\alpha_s\xi$ terms coupled with the scale choice
helping at small $x$. The value of the LO coupling is set at 
$\alpha_s(M_Z^2)=0.116$, where this LO value is unambiguous,
contrary to the normal case at LO, because the scale choice has been
determined unambiguously. The effects of varying the coupling remain
to be investigated. A standard NLO in $\alpha_s$ fit with BLM inspired scale 
fixing has also recently been performed \ref\dick{R.G. Roberts,
{\tt hep-ph/9904317}.} with less impressive results, 
particularly at small $x$. 

\medskip   

\noindent 2. Since the procedure for calculating the evolution is very
different from the NLO in $\alpha_s(Q^2)$ approach, predictions resulting
from the best fit are significantly altered. For example, the
additional terms in powers of $\bar\alpha_s\xi$ in $p^0_{2L}(\bar\alpha_s\xi)$
compared to the NLO in $\alpha_s(Q^2)$ approach more than compensates for the
decrease in the effective coupling at moderate $x$ and $Q^2$, leading
to a smaller $F_L(x,Q^2)$ (very similar to that predicted in \lorsc\
if $Q^2\geq 15 \Gev^2$)  being required to obtain a similar rate of
evolution for $F_2(x,Q^2)$. Predictions for other processes,
e.g. Drell-Yan production, are potentially very different in the two
approaches.

\medskip

\noindent 3. There is a failure of the NLO in $\alpha_s(Q^2)$ approach
at small $x$ for $Q^2\leq 2-3\Gev^2$. This can be seen in two ways. If
the gluon (and hence $F_L(x,Q^2))$ is required to be positive definite
down to $Q^2<1\Gev^2$ then the value of $(\partial F_2(x,Q^2)/\partial\ln
Q^2)$ becomes too large for $Q^2\leq 2-3\Gev^2$ \ref\caldwell{A. Caldwell
, talk at the DESY Theory Workshop on ``Recent Developments in QCD'',
October 1997 (unpublished).} (a plot can be found in \ref\zeusfit{ZEUS
collaboration: J. Breitweg {\it et al.},
{\it Eur. Jour. Phys.} \vyp{C7}{1999}{609}.}), as can be seen by
comparing the data with the prediction from a GRV type 
parameterization \ref\grv{M. Gl\"uck, 
E. Reya and A. Vogt, \ZP \vyp{C67}{1995}{433}; {\it Eur. Jour. Phys.} 
\vyp{C5}{1998}{461}.}. Alternatively,
the value of $(\partial F_2(x,Q^2)/\partial\ln Q^2)$ can be made correct down
to $\sim 1 \Gev^2$, at the expense of having a valencelike
gluon distribution, and hence odd shaped $F_L(x,Q^2)$ (see below), at
$Q^2 = 1\Gev^2$, and hence negative gluon and $F_L(x,Q^2)$ below this 
\mrst\zeusfit. Each case demonstrates that the NLO in $\alpha_s(Q^2)$
approach is breaking down at $Q^2\sim 2-3\Gev^2$ at small $x$.\foot{I
note that despite reports to the contrary an analysis of data using the
leading $\ln(1/x)$ terms with $\alpha_s(Q^2)$ does not fail in any
more dramatic a manner than this. As shown in \thornedis, using the
LO physical anomalous dimensions to perform the analysis, rather than
some factorization scheme which leads to extremely ambiguous results
at small $x$, a fit of even better quality than the NLO in
$\alpha_s(Q^2)$ fit can be achieved. The only failings are that the
pathological behaviour in the predicted $F_L(x,Q^2)$ sets in at very
slightly higher $Q^2$, and of course the NLO corrections using this
approach appear to be huge.} While
this might not seem surprizing since there are many potential reasons
for this failure (higher twist, higher orders and of course $\ln(1/x)$ 
resummations), it is a problem not shared by the full LORSC fit with the
correct scale (even though it is a considerably better fit 
at small $x$ than in \mrst). 
Because the small $x$ effective coupling becomes
proportionally smaller compared to $\alpha_s(Q^2)$ as we tend to lower
$Q^2$, and because, as seen in table 1, the coefficients in the expansion
of $p^0_{LL}(\bar\alpha_s\xi)$ are small, the evolution of
$F_L(x,Q^2)$ is slowed down at very small $x$ and $Q^2$ compared to the NLO
in $\alpha_s(Q^2)$ approach. Hence, the $F_L(x,Q^2)$ predicted by the
global fit does not evolve backwards into a pathological form at 
$Q^2=1\Gev^2$. This is shown in \fig\fllowq{Comparison of the
predictions for $F_L(x,Q^2)$ at $Q^2=1.2\Gev^2$ from the global fit
performed in this paper and the NLO in $\alpha_s(Q^2)$ fit in \mrst.}  
where I compare the predicted 
$F_L(x,Q^2)$ with that obtained from the MRST analysis at
$Q^2=1.2\Gev^2$. Clearly the shape of the LORSC $F_L(x,Q^2)$ is not
dissimilar to that of $F_2(x,Q^2)$ at the same $Q^2$, while the MRST
$F_L(x,Q^2)$ is rather odd, though it looks sensible by about
$2\Gev^2$.  (The rise at very small $x$ in the MRST curve is due to the
small quark contribution becoming dominant over the large but
valencelike gluon contribution.) Evolving downwards the MRST
$F_L(x,Q^2)$ dips down to negative values at about $1\Gev^2$ while the
LORSC $F_L(x,Q^2)$ will clearly be sensible to much lower values (this
will be investigated in detail in \thornei)). Since the effective 
coupling at small
$x$ is so small it seems reasonable to believe that the full LORSC
calculation
should really represent the physics down to low $Q^2$, as it does,
whereas even if the NLO in $\alpha_s(Q^2)$ approach had worked we
would not have known why.\foot{A recent discussion of the ``Caldwell
plot'' using the LO BFKL equation with running coupling, 
though with very different
techniques from those used in this paper, appears in 
\ref\zoller{N.N. Nikolaev and
V.R. Zoller, {\it JETP Lett.} \vyp{69}{1999}{103}.}}

\medskip

Hence, all details of the phenomenology of the scale fixed LORSC
analysis seem very satisfactory, being a distinct improvement on the
standard approach and the LORSC analysis with $Q^2=\mu^2$. As a word
of caution, the analysis presented is still a little approximate, and
all quantitative results are likewise approximate. A more careful
detailed analysis will appear soon, though it would be very surprizing
if the same quality fit were not achieved simply by a slight
alteration of input parameters and hence very slightly different
predictions.   
 
\newsec{Conclusions.}

I have presented a full discussion of the effect of the NLO
corrections to the BFKL equation. I have shown that if one resums the 
$\ln(k^2/\mu^2)$ terms into a running coupling constant, as must be
roughly correct, this alters the whole structure of
the solution to the  BFKL equation. As previously pointed out 
\janpcoll\ciafrun, at leading
twist it leads to the solution factorizing into a input dependent part
which requires regularization, i.e. is infrared renormalon contaminated,
and a $k^2$-dependent part which is well defined. The degree of
uncertainty associated with the input part is shown to have exactly the
behaviour predicted by Mueller \mueller\mplusk. 
However, this ambiguity affects the input part only, not the whole solution. 
I note that the evolution part as a function of $\gamma$
and $N$ no longer has singularities to the right of zero for
either $\gamma$ or $N$, a result which has previously been noted
\janpcoll\haakman, but seems to have been universally ignored. 
Hence, this calculable $k^2$-dependent solution has
no true powerlike behaviour in either $k^2$ or $x$ - the hard pomeron
intercept is zero. These results require no assumptions at all. If one
takes the running of the coupling in the BFKL equation seriously, the
input term is indeterminate unless $13\beta_0^2\alpha^3_s(Q_0^2)\xi\ll
1$, and the evolution term is well defined and calculable, and has no
true powerlike behaviour. 
This is not difficult to understand in a
qualitative manner. It has long been known that the typical virtuality
of a gluon in the ladder representing the BFKL Green's function has
a mean of order $k^2$, but a deviation of order
$(\bar\alpha_s\xi)^{\half}$ \diff. I have shown that the diffusion of $k^2$
into the infrared influences only the input dependent solution, the
strong coupling then leading to infrared renormalons, while the 
$k^2$-dependent part is influenced only by the ultraviolet diffusion. This
means that as one goes to smaller and smaller $x$ the appropriate
scale becomes larger and larger, the coupling weaker and weaker 
(like $\xi^{-\half}$), and
the growth from the $\ln(1/x)$ terms is sufficiently weakened by the
coupling to destroy the powerlike behaviour. 

Using the LO BFKL solution with running coupling
I have argued that in order to investigate perturbatively calculable
physics one must investigate physical anomalous dimensions \cat, or
splitting functions, which tell one how unambiguous physical
quantities evolve in terms of each other, and hence are themselves 
unambiguous, i.e. independent of factorization schemes or scales. This
is important when using a small $x$ expansion even at low orders due
to large factorization scheme uncertainties, but is now vital in
order to obtain well-defined, perturbatively calculable results. While,
of course, it is ultimately necessary to use real structure functions
$F_2(x,Q^2)$ and $F_L(x,Q^2)$, one may for simplicity work with an
unphysical, but unambiguously defined gluon structure function $G(x,Q^2)$. 
By calculating the solutions for the $Q^2$-dependent factors 
of the structure functions about the
saddle points, one obtains ordered power series in
$\beta_0\alpha_s(Q^2)$ for the physical anomalous dimensions. 
While these series appear to be very badly
convergent, the coefficients oscillate in sign, rendering them summable. I
hypothesize that one can approximate the whole result by using the BLM
scale fixing procedure \BLM\ absorbing the NLO $\beta_0$-dependent
term into the definition of the scale used in the LO expression. 
This results in an effective coupling of the form 
$1/(\beta_0(\ln(Q^2/\Lambda^2)+3.63(\bar\alpha_s(Q^2)\xi)^{\half}))$
as $x\to 0$. For different physical variables the moderate
$x$ couplings are slightly different but the asymptotic form is universal.
It is not guaranteed that this choice of coupling is really correct. 
However, the explicit NNLO calculation supports the procedure
strongly, and it is also consistent with the qualitative features one
knows must be associated with the full summation, i.e. it smooths out
the powerlike growth in $x$ in precisely the correct manner, as well
as the picture of ultraviolet diffusion. 

Examining the full NLO BFKL equation I find that as far as running
coupling is concerned by far the dominant effect is produced
solely by the $\ln(k^2/\mu^2)$ term. All additional NLO 
$\beta_0$-dependent corrections lead
to modifications to the physical splitting functions which are not only 
numerically small, but are reduced by a factor of $(\bar\alpha_s(Q^2)
\xi)^{-\half}$. This indicates that it is likely that at all orders the
$\ln(k^2/\mu^2)$ terms will lead to the dominant small $x$ effects due
to running of the coupling. Indeed, at NLO the contribution to the
physical splitting function from this term is also dominant to the
conformal corrections by $(\bar\alpha_s(Q^2)
\xi)^{\half}$. The latter are of the form expected from a
renormalization group argument, i.e. a factor of $\alpha_s(\bar\alpha_s
\xi)$ up on the LO expression, while the running coupling effect is of an
unexpected, more leading form, and essentially demands to be resummed.
I also proved that if one assumes the dominance of the $\ln(k^2/\mu^2)$ terms
the appropriate scale to use at NLO is precisely the same 
as the LO scale as $x \to 0$ - 
a result which was by no means guaranteed to be true
and seems strongly suggestive of the correctness of the approach. 
It also implies that perturbation theory at small $x$ should be 
particularly convergent.  
Using this effective scale choice in the coupling I find that the remaining, 
conformal NLO
corrections to the physical anomalous dimensions are much more
under control than for the scale choice $Q^2=\mu^2$ due to the
smallness of the effective coupling as $x\to 0$. At all $x$ and $Q^2$ 
they are subdominant to the LO result, although they 
can be significant, and in the
region of $x$ and $Q^2$ probed at HERA they are numerically quite small.      

An analysis of data using the full LO physical splitting functions
containing both leading in $\ln(1/x)$ terms and all 
${\cal O}(\alpha_s)$ terms, with scale fixing appropriate to this
combined expansion scheme, is very successful. It produces a far better fit
to data  than conventional approaches, and also predicts an
$F_L(x,Q^2)$ of the same shape as $F_2(x,Q^2)$ down to $Q^2=1\Gev^2$,
and possibly below. In fact, it seems to work perfectly over the whole
range of parameter space one might hope. The fit to $F_2(x,Q^2)$ also
leads to predictions for other quantities such as $F_L(x,Q^2)$
(difficult to measure), $F^c_2(x,Q^2)$ (not much different from the
standard approach) and Drell-Yan production (if the necessary BFKL
coefficient functions were calculated).   

Since the coupling at small $x$ is weak, seemingly at all orders,
one may be optimistic that
it is possible to use even LO perturbation theory down to very low
$Q^2$ at small $x$. Indeed, the prediction is that the corrections
at NNLO and beyond will be insignificant due to the fall of the
coupling overwhelming all possible enhancement due to small $x$ terms.  
However, there are still potentially important higher twist ($\Lambda^2/Q^2$)
contributions. Nevertheless, the weakness of the coupling may make one
hope that the small $x$ higher twist effects are strongly suppressed,
for example a weaker coupling would certainly delay the onset of such
effects as shadowing \glr\ rather significantly. Also, I note that within
the small $x$ expansion there are no infrared renormalons in the
calculation of the physical anomalous dimensions. Since renormalons
lead to 
ambiguities which must be cancelled by higher twist ambiguities they
are normally taken to be estimates of the size of these higher twist 
contributions - indeed, the scale fixing for nonsinglet evolution at
high $x$ \wong\ does imply renormalons of the type already calculated
\ref\webber{Yu.L. Dokshitzer and B.R. Webber, \PL \vyp{B352}{1995}{451}\semi
M. Dasgupta and B.R. Webber, \PL \vyp{B382}{1996}{96}.}. 
The absence of the renormalons at small $x$ makes
the author at least optimistic about the smallness of higher twist effects.
Some small $x$ higher twist calculations have already been
performed \ref\bontus{C. Bontus, talk at 
Durham HERA workshop 1998, unpublished.}. 
However, since the full physical picture
at leading twist only appears when performing a full resummed
$\ln(1/x)$ calculation including running coupling effects, a true picture 
of the higher twist contributions may sadly require similar sophistication
(if this is possible). I certainly feel that any renormalon calculations 
performed at fixed order in $\alpha_s$ may not be representative of
the true small $x$ higher twist contributions. 
If the full LO, with resummed terms and scale
fixing, analysis is indeed successful to very low $Q^2$ I would regard
this as empirical evidence, if by no means a proof, of the smallness
of higher twist corrections at small $x$. 

I have commented on other approaches to the NLO BFKL equation
throughout this paper. There have also recently been alternative attempts to 
improve the apparent bad convergence of the perturbative series which are
somewhat orthogonal to the line taken in this paper. In \gavin\ and 
\ciafpcol\ progress is made by finding resummations which improve the
convergence of the expansion of the kernel, thus implying a sensible, stable
pomeron intercept. I have no argument with this approach 
and believe that for single scale processes it is vital for obtaining
a stable expansion for general values of $x$. However, I also believe
that for structure functions it
leads to effects that are completely subdominant  to those induced by
the running of the coupling. If my assumption about the running coupling in the
kernel being accounted for by the effective $x$-dependent coupling 
in physical structure functions has any truth in it, it
makes resummations of the conformal part of the kernel unimportant
since the higher orders are so greatly weakened by the reduction in
the coupling. Hence, while the work in \gavin\ciafpcol\ is certainly 
interesting, I
believe it may be unimportant for the real physical results, at least
as far as structure functions are concerned. 

Also, there has very recently been a proposal to adopt the BLM scale
fixing procedure at the level of the {\it eigenvalues} of the kernel 
\ref\bfl{S.J. Brodsky, V.S. Fadin, V.T. Kim, L.N. Lipatov and G.B. Pivovarov,
SLAC-PUB-8037, IITAP-98-010, {\tt hep-ph/9801229}.}. 
This is similar, though not identical to the proposal for the change
in coupling proposed in \ref\ciafscale{G. Camici and M. Ciafaloni, \NP
\vyp{B496}{1997}{305}.} when the $N_f$-dependent
corrections to the NLO kernel were known.
It avoids all the running coupling effects I consider in \S 4, picking up
only those in $\gamma^1$ in \S 6, i.e. the
$\half\beta_0(\chi^2(\gamma)+\chi'(\gamma))$ and
$-(5/3)\beta_0\chi(\gamma)$ terms. This leads to a 
scale change $\ln(Q^2/\Lambda^2) \to \ln(Q^2/\Lambda^2) +A$,
where $A$ is very small (and negative). 
However, the NLO contribution to the kernel is
renormalization scheme dependent, and this result is in 
$\overline{\hbox{\rm MS}}$ scheme. By transferring to schemes that the
authors reasonably argue are more suited to gluon dominated processes,
i.e. the MOM \ref\mom{W. Celmaster and R.J. Gonsalves, \PR 
\vyp{D20}{1979}{1420} ; \PRL \vyp{42}{1979}{1435} \semi P. Pascual and 
R. Tarrach, \NP \vyp{B174}{1980}{123} ; (E) \NP \vyp{B181}{1981}{546}.} 
or $\Upsilon \to ggg$ \BLM\ schemes, the scale change at
$\gamma=\half$ becomes $\tilde Q^2  \sim 120Q^2$, and the intercept
becomes $\sim \lambda(\tilde Q^2)(1-4\alpha_s(\tilde Q^2))$. Hence, the
large increase in scale and significant reduction in the NLO
coefficient leads to a sensible NLO intercept of $\sim 0.15$ which is
not too sensitive to $Q^2$. I believe the {\it eigenvalue} of the
kernel is an inappropriate
place to make the scale choice since, as soon as one introduces the
running coupling into the BFKL equation, the whole structure changes.
The $Q^2$-dependent {\it eigenvalue} is no longer a real 
eigenvalue, as it is at strictly LO, and it
no longer has a direct physical interpretation. This is identical to
the statement that the argument of the exponent in \nlosol\ does not in
fact truly represent the full evolution of any physical quantity, is
by no means a true anomalous dimension, and should not be used for
setting the scale.  
In essence the choice in \bfl\ misses the most important
results generated by solving the BFKL equation with running coupling 
and looking at physical
quantities. This is easily seen by the fact that in any
renormalization scheme the change in scale using the method in \bfl\ 
is always of the form $\ln (\tilde Q^2/\Lambda^2) = \ln(Q^2/\Lambda^2)
+A_{rs}$, 
where $A_{rs}$ is a constant depending on the scheme.  
Using the BLM method for physical quantities, as in this paper, 
always results in $\ln(\tilde Q^2/\Lambda^2) =\ln(Q^2/\Lambda^2) +B_{rs,i}+
3.63(\bar\alpha_s(Q^2)\xi)^{\half}$, where $B_{rs,i}$ depends on
renormalization scheme and process. Clearly the $\xi$-dependent term
is the dominant one at small enough $x$ and contains the most
important physics contributing to the scale fixing. Note that this
contribution is also scheme-independent and the same at NLO as at LO, 
and that the choice of renormalization scheme
only leads to subleading contributions to the scale at small $x$.
Nevertheless, the type of renormalization scheme considered in
\bfl\ leads to a value of $B_{rs,i}$ that is rather large. This implies
that the details of calculations of structure functions in the current
experimental range may be
sensitive to the renormalization scheme chosen. However, when doing
a full analysis one should use the same scheme for all physical
splitting functions, which will be influenced by both gluon and quark
dominated processes. There are also further changes to the scale due
to the running coupling effects at ${\cal O}(\alpha_s^2)$, which
will be scheme dependent, and potentially of similar importance to
the differences in $B_{rs,i}$ at the relatively high $x$ values where
it is relevant. A full understanding of the relevance of
renormalization scheme changes needs to take these into account
carefully. 

\medskip

Hence, to summarize, I believe that the method of solving for physical
quantities using the BFKL equation with running coupling and full NLO
contributions presented in this paper is the best way to proceed for
the analysis of deep inelastic scattering at small $x$. Certainly, the
conclusion that the running coupling serious alters our picture of
BFKL physics, destroying predictivity for the input and maintaining
it, but smoothing out the powerlike behaviour for the calculable
evolution, seems to be incontrovertible. More controversial
is the proposal that the true physics may be well described by a
coupling which falls as $x$ falls like $\ln(1/x)^{-\half}$. This 
is strongly supported by current finite order expansions,
the universality between DIS processes and different orders,
the diffusion picture, and the general
features that the full solution must exhibit. However, it may well be
possible to validate this more strongly, or invalidate it. Also, the
discussion in this paper has very firmly used the assumption that the
lower end of the gluon ladder is fixed at some low scale, as is
appropriate for deep inelastic scattering. Further investigation is
required in order to consider different types of process, 
although I imagine that the
qualitative results will be the same. Overall normalization will be
infrared renormalon contaminated, 
since even if there are no small scales in the problem the diffusion
into the infrared will eventually be important for small enough $x$,  
while evolution will be calculable but
not truly powerlike. If the general results of this paper are correct,
perturbative calculations at small $x$ will be very reliable and
convergent. They would also
explain why perturbation theory appears, at least qualitatively, to
be working at very low scales at small $x$, but also implies that the 
standard NLO
in $\alpha_s(Q^2)$ approach is not really quantitatively correct at
small $x$. More phenomenological work, including
calculation of currently unknown coefficient functions as power series
in $\alpha_s\ln(1/x)$, would then be important in order to produce truly
precise calculations for small $x$ physics.

\bigskip

\noindent{\bf Acknowledgements.}
\medskip
I would like to thank Dick Roberts for continual help during the period of 
this work and for the use of the MRS fit program. I would also
like to thank Stefano Catani, Jeff Forshaw, Alex Kovner, Martin
McDermott, Graham Ross,  Davis
Soper, George Sterman, and Mark Wusthoff for useful discussions.

\vfill 
\eject

\noindent {\bf Table 1.} \hfil\break

The coefficients in the power series $p^i_{LL}(\bar\alpha_s(Q^2)\xi)=
\sum_{0}^{\infty}a_n
(\bar\alpha_s(Q^2)\xi)^{n}/ n!$ for the
various LO and NLO contributions to the physical splitting function 
$P_{LL}(x,Q^2)$. 

\hfil\vtop{{\offinterlineskip
\halign{ \strut\tabskip=0.6pc
\vrule#&  #\hfil&  \vrule#&  \hfil#& \vrule#& \hfil#& \vrule#& \hfil#&
\vrule#& \hfil#& \vrule#\tabskip=0pt\cr
\noalign{\hrule}
& $n$ && $p^0_{LL}$ && $p^{1,tot}_{LL}$ && $p^{1,\beta}_{LL}$ &&
$p^{1,conf}_{LL}$ &\cr
\noalign{\hrule}
& 0 && 1.00 && 0.23 && -2.00 && 1.57 &\cr
& 1 && 0.00 && 4.38 && 4.15 && 1.60 &\cr
& 2 && 0.00 && 15.87 && 11.32 && 8.29 &\cr
& 3 && 2.40 && 13.41 && -16.18 && 24.25 &\cr
& 4 && 0.00 && 86.26 && 76.03 && 35.31 &\cr
& 5 && 2.07 && 252.92 && 167.34 && 140.81 &\cr
& 6 && 17.34 && 323.08 && -81.51 && 377.69 &\cr
& 7 && 2.01 && 1699.65 && 1472.42 && 713.25 &\cr
& 8 && 39.89 && 4338.69 && 2665.07 && 2553.16 &\cr
& 9 && 168.75 && 7592.65 && 1674.16 && 6470.97 &\cr
& 10 && 69.99 && 33409.13 && 28319.16 && 14435.29 &\cr
& 11 && 661.25 && 79427.26 && 47284.56 && 47746.61 &\cr
& 12 && 1945.31 && 173361.43 && 81792.97 && 118560.14 &\cr
& 13 && 1717.68 && 657395.79 && 543255.72 && 293414.46 &\cr
& 14 && 10643.26 && 1527235.16 && 927749.64 && 905642.90 &\cr
& 15 && 25266.78 && 3833618.50  && 23539999.61 && 2256438.84 &\cr
\noalign{\hrule}}}}\hfil

\vfil
\eject

\noindent {\bf Table 2.} \hfil\break

The coefficients in the power series $p^i_{2L}(\bar\alpha_s(Q^2)\xi)=
\sum_{0}^{\infty}a_n
(\bar\alpha_s(Q^2)\xi)^{n}/ n!$ for the
LO and $\beta_0$-dependent NLO contributions to the physical splitting
function $P_{2L}(x,Q^2)$. 

\hfil\vtop{{\offinterlineskip
\halign{ \strut\tabskip=0.6pc
\vrule#&  #\hfil&  \vrule#& \hfil#&
\vrule#& \hfil#& \vrule#\tabskip=0pt\cr
\noalign{\hrule}
& $n$ && $p^0_{2L}$ && $p^{1,\beta}_{2L}$ &\cr
\noalign{\hrule}
& 0 && 2.50 && -4.00 &\cr
& 1 && 1.00 && 9.39 &\cr
& 2 && 1.00 && 36.60 &\cr
& 3 && 7.01 && 6.27 &\cr
& 4 && 5.81 && 239.73 &\cr
& 5 && 13.40 && 687.03 &\cr
& 6 && 58.11 && 771.35 &\cr
& 7 && 64.74 && 5281.50 &\cr
& 8 && 196.83 && 13213.51 &\cr
& 9 && 649.89 && 24043.80 &\cr
& 10 && 930.65 && 111578.92 &\cr
& 11 && 3034.70 && 265509.09 &\cr
& 12 && 8527.87 && 613964.05 &\cr
& 13 && 15046.02 && 2311855.03 &\cr
& 14 && 48434.53 && 5521425.31 &\cr
& 15 && 124600.51 && 14458201.96 &\cr
\noalign{\hrule}}}}\hfil

\vfil
\eject

\noindent Table 3\hfil\break
\noindent Comparison of quality of fits using full leading order (including 
$\ln (1/x)$ terms) renormalization scheme consistent expression, with BLM 
scale setting and the NLO in $\alpha_s(Q^2)$ fit \mrst. The references to the 
data can be found in \mrst.   
\medskip

\hfil\vtop{{\offinterlineskip
\halign{ \strut\tabskip=0.6pc
\vrule#&  #\hfil&  \vrule#&  \hfil#& \vrule#& \hfil#& \vrule#& \hfil#&
\vrule#\tabskip=0pt\cr
\noalign{\hrule}
& Experiment && data &&$\chi^2$&\omit& \omit &\cr
&\omit&& points && LO(x) &\omit& MRST &\cr
\noalign{\hrule}
& H1 $F^{ep}_2$ && 221 && 149 && 164 &\cr
& ZEUS $F^{ep}_2$ && 204 && 246 && 270 &\cr
\noalign{\hrule}
& BCDMS $F^{\mu p}_2$ && 174 && 241 && 249 &\cr
& NMC $F^{\mu p}_2$ && 130 && 118 && 141 &\cr
& NMC $F^{\mu d}_2$ && 130 && 81 && 101 &\cr
& NMC $F^{\mu n}_2/F^{\mu p}_2$ && 163 && 176 && 187 &\cr
& SLAC $F^{\mu p}_2$ && 70 && 87 && 119 &\cr
& E665 $F^{\mu p}_2$ && 53 && 59 && 58 &\cr
& E665 $F^{\mu d}_2$ && 53 && 61 && 61 &\cr
\noalign{\hrule}
& CCFR $F^{\nu N}_2$ && 66 && 57 && 93 &\cr
& CCFR $F^{\nu N}_3$ && 66 && 65 && 68 &\cr
\noalign{\hrule}
& total && 1330 && 1339 && 1511 &\cr
\noalign{\hrule}}}}\hfil

\vfill\eject\immediate\closeout\ffile{\parindent40pt
\baselineskip14pt\centerline{{\bf Figure Captions}}\nobreak\medskip
\escapechar=` \input figs.tmp\vfill\eject}

\footatend\vfill\supereject\immediate\closeout\rfile\writestoppt
\baselineskip=14pt\centerline{{\bf References}}\bigskip{\frenchspacing%
\parindent=20pt\escapechar=` \input refs.tmp\vfill\eject}\nonfrenchspacing

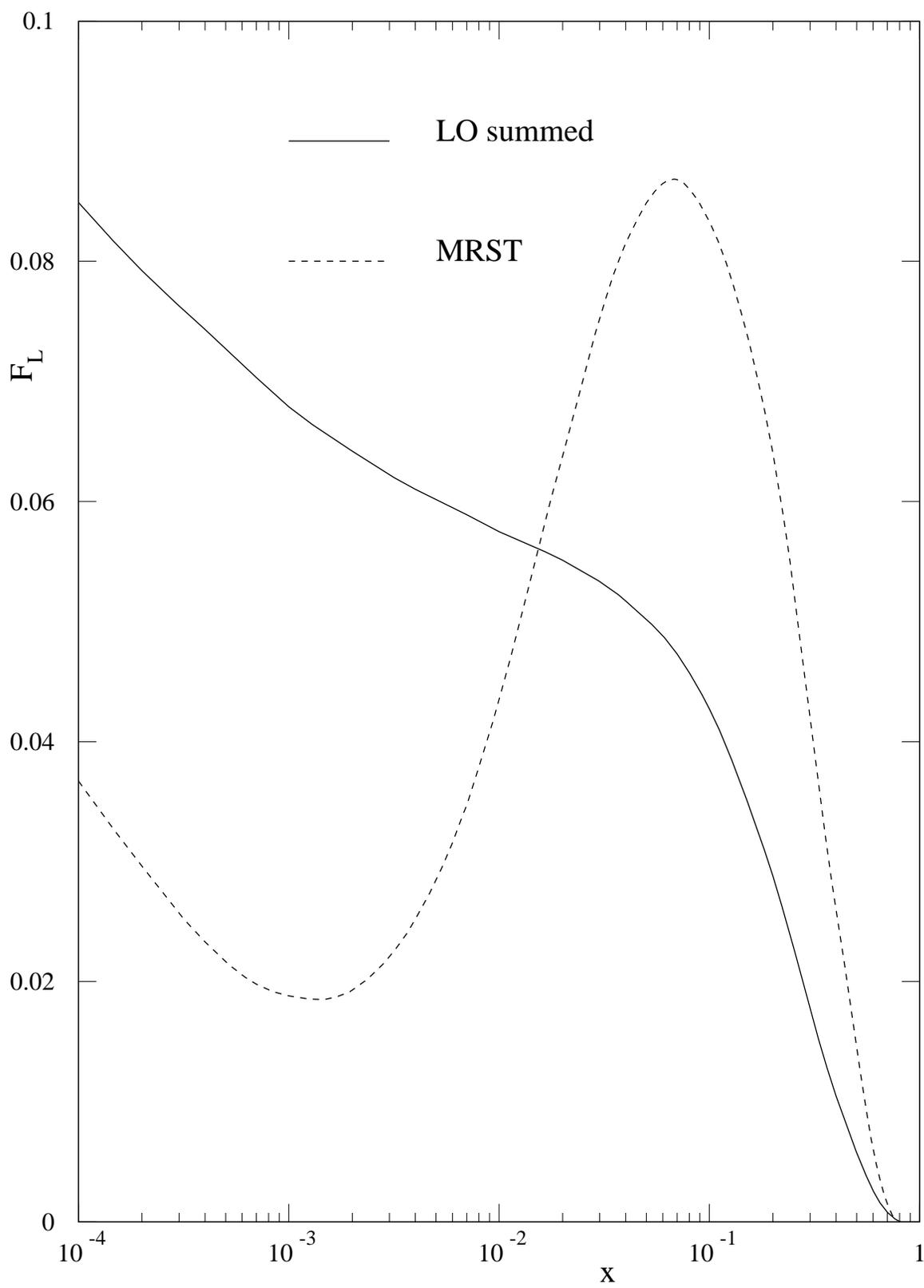
\end